\documentclass[journal]{IEEEtran}
\IEEEoverridecommandlockouts
\makeatletter
\def\endthebibliography{%
  \def\@noitemerr{\@latex@warning{Empty `thebibliography' environment}}%
  \endlist
}
\makeatother

\usepackage{cite}
\usepackage{amsmath,amssymb,amsfonts}
\usepackage{algorithm}
\usepackage{algpseudocode}
\usepackage{graphicx}
\usepackage{textcomp}
\usepackage[most]{tcolorbox}
\usepackage{xcolor}
\usepackage{balance}
\usepackage{multirow}
\usepackage{booktabs}
\usepackage{url}
\usepackage[caption=false]{subfig}
\usepackage[belowskip=-8pt,aboveskip=2pt]{caption}
\usepackage{float}
\usepackage{stfloats}
\usepackage{enumitem}
\usepackage[colorlinks,
            linkcolor=black,
            anchorcolor=black,
            citecolor=black]{hyperref}
\pdfoutput=1

\newcommand{\wap}{\textsc{WirelessAgent\textsuperscript{++}}}
\newcommand{\wb}{\textsc{WirelessBench}}

\newtheorem{definition}{Definition}
\newtheorem{remark}{Remark}
\newtheorem{problem}{Problem}

\ifodd 1

\newcommand{\com}[1]{{\color{blue}#1}}
\else

\newcommand{\com}[1]{}
\fi

\def\BibTeX{{\rm B\kern-.05em{\sc i\kern-.025em b}\kern-.08em
    T\kern-.1667em\lower.7ex\hbox{E}\kern-.125emX}}

\begin{document}

\title{WirelessAgent++: Automated Agentic Workflow Design and Benchmarking for Wireless Networks}

\author{Jingwen Tong,~\IEEEmembership{Member,~IEEE},  Zijian Li, Fang Liu, Wei Guo,~\IEEEmembership{Member,~IEEE}, and Jun Zhang,~\IEEEmembership{Fellow,~IEEE}
\thanks{Jingwen Tong and Fang Liu are with the College of Electronics and Information Engineering, Shenzhen University, Shenzhen, China (e-mails: \{eejwentong, liuf\}@szu.edu.cn);
Zijian Li, Wei Guo, and Jun Zhang are with the Department of Electronic and Computer Engineering, The Hong Kong University of Science and Technology, Hong Kong SAR (e-mails: zijian.li@connect.ust.hk; eeweiguo@ust.hk; eejzhang@ust.hk). The corresponding author is Jun Zhang.}
}

\maketitle

\begin{abstract}
The integration of large language models (LLMs) into wireless networks has sparked growing interest in building autonomous AI agents for wireless tasks. However, existing approaches rely heavily on manually crafted prompts and static agentic workflows, a process that is labor-intensive, unscalable, and often suboptimal. In this paper, we propose \wap{}, a framework that automates the design of agentic workflows for various wireless tasks. By treating each workflow as an executable code composed of modular operators, \wap{} casts agent design as a program search problem and solves it with a domain-adapted Monte Carlo Tree Search (MCTS) algorithm. Moreover, we establish \wb{}, a standardized multi-dimensional benchmark suite comprising Wireless Communication Homework (WCHW), Network Slicing (WCNS), and Mobile Service Assurance (WCMSA), covering knowledge reasoning, code-augmented tool use, and multi-step decision-making. Experiments demonstrate that \wap{} autonomously discovers superior workflows, achieving test scores of $78.37\%$ (WCHW), $90.95\%$ (WCNS), and $97.07\%$ (WCMSA), with a total search cost below $\$ 5$ per task. Notably, our approach outperforms state-of-the-art prompting baselines by up to $31\%$ and general-purpose workflow optimizers by $11.1\%$, validating its effectiveness in generating robust, self-evolving wireless agents. The code is available at \url{https://github.com/jwentong/WirelessAgent-R2}.
\end{abstract}


\section{Introduction}
\label{sec:intro}

\IEEEPARstart{T}{he} growing complexity of wireless networks, spanning heterogeneous radio access, dynamic spectrum sharing, and ultra-dense deployments, has made AI-native design a cornerstone of the 6G vision~\cite{letaief2019roadmap}. Managing the diverse wireless tasks that arise in modern networks, from base-station configuration and fault diagnosis to real-time resource optimization, demands intelligence that goes well beyond traditional model-driven or data-driven approaches~\cite{singh2025ai}. While classical optimization-based solutions excel at well-defined, single-objective problems, they struggle with open-ended problem formulations, non-ideal channel models, and the multi-step reasoning that many practical wireless tasks require.

The recent development of large language models (LLMs) is driving a promising paradigm shift~\cite{brown2020language}. Their ability to understand natural-language specifications, generate structured outputs, and perform in-context reasoning makes them attractive building blocks for intelligent network management. A growing body of work has explored this direction: TeleQnA~\cite{maatouk2024teleqna} evaluates LLMs as knowledge bases for telecom question-answering; WirelessLLM~\cite{shao2024wirelessllm} aligns LLM capabilities with wireless domain knowledge through prompting and retrieval-augmented generation; NetLLM~\cite{liu2024netllm} fine-tunes LLMs for networking tasks such as viewport prediction and adaptive bitrate streaming; and TelecomGPT~\cite{zou2024telecomgpt} further fine-tunes open-source LLMs on telecom-specific corpora for standards-based reasoning. These contributions demonstrate that LLMs can internalize substantial wireless knowledge. However, directly applying an LLM to a wireless task, in a single-turn, prompt-in-answer-out fashion, often falls short when the task requires multi-step reasoning, external tool invocation (e.g., calling a path-loss calculator), or iterative error correction.

To overcome these limitations, recent research has turned to \emph{LLM-based autonomous agents}: systems that augment an LLM with external tools, memory, and structured control flows to solve complex tasks in multiple steps~\cite{xi2023rise,wang2024survey_agent}. The ReAct paradigm~\cite{yao2023react} interleaves reasoning and acting in a closed loop, enabling agents to dynamically decide what action to take next. Orchestration frameworks such as LangChain~\cite{langchain2023} further facilitate tool integration and task decomposition. In our preliminary work, WirelessAgent~\cite{tong2024wirelessagent}, we demonstrated that a ReAct-style agent equipped with domain-specific tools (\textit{e.g.}, telecom formula retriever, precision calculator, and ray-tracing channel predictor) can perform multi-step telecom reasoning and achieve near-optimal performance on network slicing tasks. This established a proof of concept: agentic workflows are a viable and powerful approach to wireless AI.

However, these approaches suffer from a fundamental limitation: \emph{the agentic workflow is entirely hand-crafted and static}. Researchers manually decide which tools to invoke, when to reflect on intermediate results, how to decompose a problem into sub-tasks, and how to format the final output. This ``manual agent engineering'' is becoming the new bottleneck for wireless AI, just as manual feature engineering was the bottleneck for machine learning before the advent of representation learning~\cite{hu2024adas}. As tasks diversify and models scale, the human effort required to design, tune, and maintain these workflows grows prohibitively, and the resulting designs are often suboptimal for the task at hand. This bottleneck calls for automated agentic workflow design in wireless networks.

Automating the design of agentic workflows for wireless networks presents several challenges.  First, the search space is vast and code-structured, where each workflow is an executable program with diverse control flow, prompt strings, and operator compositions. Second, wireless tasks require domain-specific tools (\textit{e.g.,} ray-tracing predictors, Kalman filters, precision calculators) that must be seamlessly integrated into the agent's reasoning loop, not merely invoked as static function calls. Third, the evaluation is inherently noisy and costly, as each candidate workflow must be executed on representative problems involving multiple LLM application programming interface (API) calls.

In this paper, we propose \wap{}, a framework that casts agent design as a \emph{program search problem}, which is analogous to neural architecture search (NAS)~\cite{zoph2017nas} but operating over agent workflow topologies. We devise a domain-adapted Monte Carlo Tree Search (MCTS) method to solve this program search problem by jointly optimizing the workflow and tool calling. In addition, we introduce a \emph{ReAct-based ToolAgent} operator that enables closed-loop interleaving of LLM reasoning and external tool calls. Moreover, we employ three domain-aware enhancements (\textit{i.e.}, penalized Boltzmann selection, maturity-aware heuristic critic, and 3-class experience replay) to address the evaluation noise, cost, and task diversity challenges unique to the wireless domain. To enable rigorous and reproducible evaluation, we further establish\wb{}, a standardized multi-dimensional benchmark suite comprising Wireless Communication Homework (WCHW), Network Slicing (WCNS), and Mobile Service Assurance (WCMSA), covering knowledge reasoning, code-augmented tool use, and multi-step decision-making.

We conduct extensive experiments on \wb{} to evaluate \wap{}. Numerical results demonstrate that \wap{} autonomously discovers workflows achieving test scores of $78.37\%$ (WCHW), $90.95\%$ (WCNS), and $97.07\%$ (WCMSA). Notably, these automatically generated workflows outperform state-of-the-art hand-crafted prompting baselines by up to $31\%$ and the best general-purpose workflow optimizer by $11.1\%$, validating both the feasibility and the practical value of automated agent design in the wireless domain. The main contributions of this work are summarized as follows:
\begin{itemize}[leftmargin=*]
    \item We formalize \emph{automated agent design for wireless networks} as a program search problem, representing each workflow as executable code composed of modular operators and solving it with a domain-adapted MCTS algorithm.

    \item We design a \emph{ReAct-based ToolAgent} operator for closed-loop tool use and propose three domain-aware enhancements, \textit{i.e.}, penalized Boltzmann selection, maturity-aware heuristic critic, and 3-class experience replay, tailored to the complex evaluation landscape of wireless tasks.

    \item We release \wb{}, a first and multi-dimensional benchmark for wireless LLM-based agents, comprising WCHW (knowledge reasoning), WCNS (dynamic network slicing), and WCMSA (proactive mobile service assurance), totaling 3,392 problems with deterministic ground truths.

    \item Extensive experiments show that automatically discovered workflows outperform hand-crafted baselines by up to $31\%$ and the best general-purpose optimizer by $11.1\%$.
\end{itemize}

The remainder of this paper is organized as follows. Section~\ref{sec:related} reviews related works. Section~\ref{sec:problem} provides a system overview and formulates the automated agent design problem. Section~\ref{sec:method} details the \wap{} framework. Section~\ref{sec:benchmark} introduces \wb{}. Section~\ref{sec:experiments} presents the experimental results. Section~\ref{sec:conclusion} concludes the paper.

\section{Related Work}
\label{sec:related}

\subsection{LLMs for Wireless Networks}
The application of LLMs to wireless networks has advanced rapidly, spanning telecom Q\&A~\cite{maatouk2024teleqna}, domain-knowledge alignment via prompting and retrieval-augmented generation~\cite{shao2024wirelessllm}, LLM fine-tuning for networking tasks~\cite{liu2024netllm}, and telecom-corpus pre-training~\cite{zou2024telecomgpt}. These efforts enhance what an LLM \emph{knows} about wireless systems, yet the agent's \emph{reasoning structure}, \textit{i.e.}, how it decomposes a problem, when to invoke tools, and how to aggregate results, remains fixed and manually specified. Our work automates this structural design itself.

\subsection{Autonomous Agentic Workflow Design}
LLM-based autonomous agents~\cite{xi2023rise,wang2024survey_agent} extend standalone LLMs with tool use and multi-step control flow. Foundational paradigms include ReAct~\cite{yao2023react} (reasoning--acting loops), CodeAct~\cite{wang2024codeact} (executable code as actions), and orchestration frameworks such as LangChain~\cite{langchain2023} and AutoGPT~\cite{autogpt2023}. However, all rely on hand-crafted workflows. Recent work seeks to automate agent design: OPRO~\cite{yang2024opro} and DSPy~\cite{khattab2023dspy} optimize prompts and pipelines; ADAS~\cite{hu2024adas} programs new agent architectures via a meta-agent; AutoFlow~\cite{li2024autoflow} iteratively refines natural-language-program workflows; AFlow~\cite{zhang2025aflow} uses MCTS over executable code; and AgentFlow~\cite{li2025agentflow} trains a four-module agent with reinforcement learning. These works, however, target general NLP benchmarks and do not address wireless-specific challenges such as noisy evaluation, high stability requirements, and tight tool integration. Our work bridges this gap by applying automated agentic workflow design to wireless tasks with a standardized wireless benchmark.

\subsection{AI Agents for Wireless Networks}
The vision of intelligent agents for wireless networks dates back to the AI-router concept of Jiang~\emph{et~al.}~\cite{jiang2020intelligent_agent}. The LLM era has reinvigorated this line of research: AgentRAN~\cite{elkael2025agentran} builds a self-organizing hierarchy of LLM agents for Open~RAN; Agoran~\cite{chatzistefanidis2025agoran} deploys cooperating AI branches in a 6G RAN marketplace; Agentic TinyML~\cite{saleh2025agentic_tinyml} places lightweight agents at edge nodes for proactive handover; Lin~\emph{et~al.}~\cite{lin2025llm_network} enable natural-language network control; SignalLLM~\cite{ke2025signalllm} constructs a general-purpose agent for signal processing; and our prior WirelessAgent~\cite{tong2024wirelessagent} demonstrated ReAct-style multi-step telecom reasoning with domain tools. Despite this progress, \emph{all} existing wireless agents rely on manually designed workflows. \wap{} is, to our knowledge, the first framework to automate agentic workflow design for wireless networks.

\section{System Overview and Problem Formulation}
\label{sec:problem}

\subsection{From WirelessAgent to \wap{}}
\label{sec:overview}

\begin{figure*}[t]
    \centering
    \includegraphics[width=2\columnwidth]{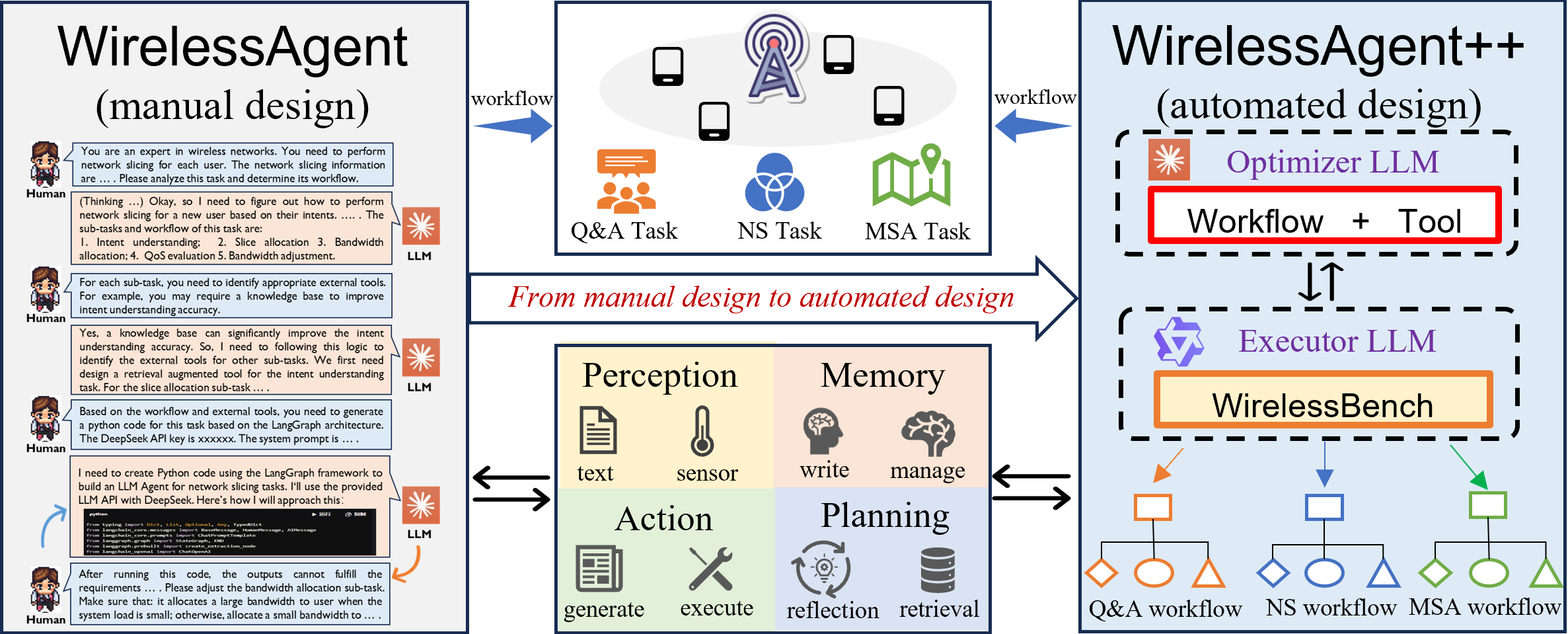}
    \caption{System overview: from WirelessAgent (manual design) to \wap{} (automated design).  Left: in WirelessAgent, a human expert iteratively designs a fixed agentic workflow through multi-round dialogue with an LLM.  Right: in \wap{}, an Optimizer LLM jointly searches over the workflow structure and tool-calling strategy; the resulting workflow is executed by an Executor LLM on \wb{}, automatically producing distinct, task-adaptive workflows (e.g., Q\&A, network slicing (NS), and mobile service assurance (MSA)) without manual engineering.}
    \label{fig:system_overview}
\end{figure*}

Our prior work, WirelessAgent~\cite{tong2024wirelessagent}, introduced an LLM-based agent framework for wireless networks built on four core modules (\textit{i.e.}, perception, memory, planning, and action) that mirror human cognitive processes.  The framework employs a ReAct-style agentic workflow~\cite{yao2023react} implemented on the LangGraph architecture, where the agent interleaves reasoning with domain-specific tool calls (\textit{e.g.,} a telecom formula retriever, a precision calculator, and a ray-tracing channel predictor) in a closed loop.  In a network slicing case study, WirelessAgent achieved near-optimal bandwidth utilization, outperforming prompt-only baselines by $44.4\%$.

Despite these encouraging results, WirelessAgent suffers from three limitations that motivate the present work.  First, the workflow topology, which tools to invoke, in what order, and how to aggregate intermediate results, are entirely \emph{hand-coded} by a human expert for each task.  Second, extending the framework to a new task (\textit{e.g.,} mobile service assurance) requires redesigning the entire workflow from scratch, as the static control flow cannot generalize across task types. Third, the prompt strings, error-handling logic, and output formatting are manually tuned without systematic optimization, leaving significant performance on the table.

As illustrated in Fig.~\ref{fig:system_overview}, \wap{} addresses these limitations by replacing the human expert with an automated program search pipeline. It automates this process: an Optimizer LLM jointly searches over workflow structures and tool-calling strategies, while an Executor LLM evaluates each candidate workflow on \wb{}.  Through MCTS-based iterative optimization, the framework automatically produces distinct, task-adaptive workflows, for various wireless tasks, such as telecom Q\&A, NS, and MSA tasks.

\subsection{Wireless Tasks as Agent Programs}
To formalize this automated search, we next define the key concepts, \textit{i.e.}, wireless tasks, agent workflows, operators, and the search space that underpin the \wap{} optimization framework.
Mathematically, we define a wireless task by $\mathcal{T} = (\mathcal{D}, \mathcal{M})$, where $\mathcal{D} = \{(x_i, y_i)\}_{i=1}^{N}$ is a dataset of question--answer pairs and $\mathcal{M}$ is an evaluation metric (e.g., accuracy, F1 score, or a composite multi-dimensional score). An \emph{agent workflow} $\mathcal{W}$ is an executable program that, given an input question $x$, produces an output $\hat{y}$ through a sequence of operations:
\begin{equation}
    \hat{y} = \mathcal{W}(x; \boldsymbol{\theta}),
\end{equation}
where $\boldsymbol{\theta}$ is the configuration parameters, including prompt strings, operator choices, and the control-flow structure.

\begin{definition}[Operator]
An \emph{operator} is an atomic building block of a workflow. We define the operator set $\mathcal{O} = \{o_1, o_2, \ldots, o_K\}$, where each operator $o_k$ has a typed interface $o_k: \mathcal{X}_k \rightarrow \mathcal{Y}_k$. Specifically, the operator library provides:
\begin{itemize}[leftmargin=*]
    \item \texttt{Custom}$(x, p)$: It invokes an LLM with input $x$ and instruction prompt $p$;
    \item \texttt{ToolAgent}$(x, s)$: A ReAct-based agent~\cite{yao2023react} that interleaves \emph{reasoning} (``what tool should I call next?'') and \emph{acting} (executing a tool call) in a loop for up to $s$ iterations, with access to domain-specific tools such as ray-tracing channel predictors and Kalman filters;
    \item \texttt{CodeLevel}$(x, f)$: A deterministic, \emph{LLM-free} operator that directly executes a domain tool $f\!\in\!\mathcal{F}$ on input $x$ via compiled code (e.g., \texttt{CodeLevelRayTracing}, \texttt{CodeLevelKalmanPredictor}). No LLM call is made; the output is exact and reproducible;
    \item \texttt{ScEnsemble}$(\{y_i\}, x)$: It aggregates multiple candidate answers via self-consistency voting;
    \item \texttt{AnswerGenerate}$(x)$: It produces a structured answer.
\end{itemize}
\end{definition}

\begin{remark}[Two Tool-Invocation Modes]
\texttt{ToolAgent} and \texttt{CodeLevel} represent two complementary ways to invoke domain tools. \texttt{ToolAgent} wraps tool calls inside an LLM reasoning loop, allowing the agent to \emph{dynamically decide} which tool to call, with what arguments, and how to interpret the result. \texttt{CodeLevel} bypasses the LLM entirely and calls the tool via a deterministic code, yielding zero-variance outputs at near-zero cost. During MCTS optimization, the search algorithm is free to choose either mode; as we show in Section~\ref{sec:case_studies}, the optimizer often discovers that an initial \texttt{ToolAgent}-based workflow can be refined into a more efficient \texttt{CodeLevel} pipeline once the correct tool-calling pattern has been identified.
\end{remark}

The detailed pseudocode for the \texttt{ToolAgent} operator is given in Algorithm~\ref{alg:react}.

\begin{algorithm}[t]
\caption{ReAct-Based ToolAgent Operator}
\label{alg:react}
\small
\begin{algorithmic}[1]
\Require Input $x$, strategy prompt $s$, tool set $\mathcal{F}$, max iterations $I$
\Ensure Output $\hat{y}$
\State $\mathcal{H} \gets [\,]$ \Comment{Interaction history}
\State $n_{\text{fail}} \gets 0$ \Comment{Consecutive failure counter}
\For{$i = 1, 2, \ldots, I$}
    \State \textbf{Think:} $(\texttt{thought}_i, a_i, \texttt{args}_i) \gets \text{LLM}(x, s, \mathcal{H}, \mathcal{F})$
    \If{$a_i = \texttt{finish}$}
        \State \Return $\texttt{args}_i$ \Comment{Agent decides to stop}
    \EndIf
    \State \textbf{Act:} $r_i \gets \mathcal{F}[a_i](\texttt{args}_i)$ \Comment{Execute tool}
    \If{$r_i$ is error}
        \State $n_{\text{fail}} \gets n_{\text{fail}} + 1$
        \If{$n_{\text{fail}} \geq 2$} \textbf{break} \Comment{Early stop}
        \EndIf
    \Else
        \State $n_{\text{fail}} \gets 0$
    \EndIf
    \State \textbf{Observe:} Append $(\texttt{thought}_i, a_i, r_i)$ to $\mathcal{H}$
\EndFor
\State $\hat{y} \gets \text{LLM}(x, \mathcal{H}[-3:])$ \Comment{Fallback: summarize last 3 steps}
\State \Return $\hat{y}$
\end{algorithmic}
\end{algorithm}

The \texttt{ToolAgent} prompt is structured in three layers: (i)~a \emph{strategy prompt} that is optimizable by the MCTS search, (ii)~a \emph{runtime context} that dynamically includes tool descriptions, the current problem, and the interaction history, and (iii)~a \emph{fixed protocol} that enforces structured output in XML format (\texttt{<thought>}, \texttt{<action>}, \texttt{<action\_input>}). Tool arguments are parsed with automatic correction for common LLM output issues such as double braces, unescaped LaTeX, and inline math expressions in JSON fields. If the maximum iteration count $I$ is reached without a \texttt{finish} action, the agent falls back to generating an answer from the last three interaction steps.

\subsection{Search Space}
The search space $\Omega$ is the set of all syntactically valid executable programs (or workflows) that:
\begin{enumerate}[leftmargin=*]
    \item Implement a \texttt{Workflow} class with an asynchronous \texttt{\_\_call\_\_} method;
    \item Use only operators from $\mathcal{O}$ and tools from the domain library $\mathcal{F}$;
    \item Return a tuple $(\hat{y}, c)$ where $\hat{y}$ is the answer string and $c$ is the API cost.
\end{enumerate}
Note that this space is vast but structured: workflows are short programs (typically 10--50 lines) expressed through a constrained API, which makes the search tractable.

\subsection{Optimization Objective}
\begin{problem}[Automated Agent Design]
Given a wireless task $\mathcal{T} = (\mathcal{D}, \mathcal{M})$, a set of operators $\mathcal{O}$, and a domain tool library $\mathcal{F}$, find the workflow $\mathcal{W}^*$ that maximizes the objective:
\begin{equation}
    \mathcal{W}^* = \arg\max_{\mathcal{W} \in \Omega} \; \mathbb{E}_{(x,y) \sim \mathcal{D}_{\mathrm{val}}} \left[ \mathcal{M}\!\left(\mathcal{W}(x), y\right) \right].
    \label{eq:objective}
\end{equation}
\end{problem}

In practice, we evaluate $\mathcal{W}$ on a validation subset $\mathcal{D}_{\mathrm{val}} \subset \mathcal{D}$ and use the median score across multiple evaluation runs for robustness, as detailed in Section~\ref{sec:method}.

\section{The\wap{} Framework}
\label{sec:method}

This section presents the \wap{} framework. We first describe the system architecture (Section~\ref{sec:arch}), then detail the domain-specific tool library (Section~\ref{sec:tools}), the core MCTS-based optimization algorithm (Section~\ref{sec:mcts}), the maturity-aware heuristic critic (Section~\ref{sec:critic}), and the 3-class experience replay mechanism (Section~\ref{sec:experience}).

\subsection{System Architecture}
\label{sec:arch}
\begin{figure*}[t]
    \centering
    \includegraphics[width=2\columnwidth]{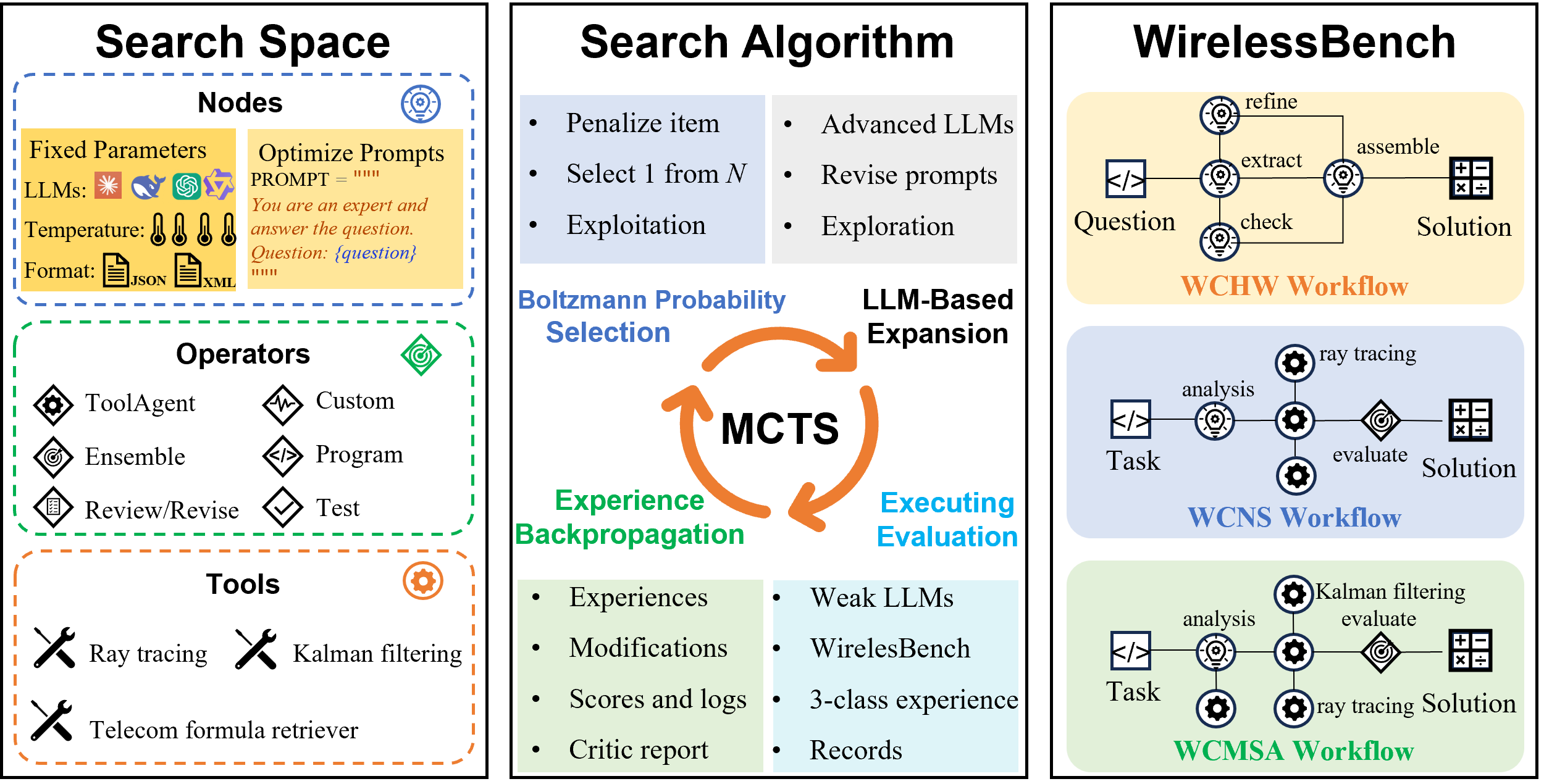}
    \caption{The overall architecture of \wap{}, comprising three components. \textbf{Search Space:} each workflow node contains fixed parameters and optimizable prompts, composed from a library of operators and domain-specific tools. \textbf{Search Algorithm:} an MCTS-based optimizer iterates through four phases: penalized Boltzmann probability selection, LLM-based expansion, executing evaluation, and experience backpropagation, using an advanced Optimizer LLM for expansion and a cost-efficient Executor LLM for evaluation. \textbf{\wb{}:} example task-adaptive workflows automatically discovered for WCHW, WCNS, and WCMSA, each exhibiting distinct operator compositions and tool-calling patterns.}
    \label{fig:architecture}
\end{figure*}

Fig.~\ref{fig:architecture} illustrates the \wap{} framework, which consists of three tightly coupled components: a structured search space, an MCTS-based search algorithm, and the \wb{} evaluation suite.

\begin{enumerate}[leftmargin=*]
    \item \textbf{Search Space (left):} Each candidate workflow is represented as a \emph{node} in the MCTS tree. A node consists of \emph{fixed parameters}, \textit{i.e.}, the LLM backbone, temperature, and output format (\textit{e.g.}, JSON, XML), and \emph{optimizable prompts} that the search algorithm revises across iterations. Nodes are composed from a library of modular operators, including \texttt{Custom} (LLM reasoning), \texttt{ToolAgent} (ReAct-style tool use~\cite{yao2023react}), \texttt{Ensemble} (self-consistency voting), \texttt{Review/Revise} (self-critique and correction), \texttt{Program} (code generation and execution), and \texttt{Test} (output validation), together with a domain-specific tool set $\mathcal{F}$ comprising a ray-tracing channel predictor, a Kalman filter, and a telecom formula retriever.

    \item \textbf{Search Algorithm (center):} The core MCTS loop iterates through four phases: (i)~\emph{Boltzmann Probability Selection} chooses a parent workflow from the top-$K$ candidates, balancing exploitation of high-scoring nodes with exploration of under-visited ones; (ii)~\emph{LLM-Based Expansion} employs an advanced Optimizer LLM (\textit{e.g.}, GPT-4o) to propose a focused code mutation, \textit{e.g.}, revising prompts, adding or removing operators, or changing tool-calling strategies; (iii)~\emph{Executing Evaluation} deploys the mutated workflow on \wb{} using a cost-efficient Executor LLM (\textit{e.g.}, DeepSeek-V3) and computes a robust median score; and (iv)~\emph{Experience Backpropagation} records the modification outcome, error logs, and scores to guide future iterations. This two-tier LLM design, \textit{i.e.}, an advanced model for creative exploration and a cost-efficient model for large-scale evaluation, keeps optimization costs manageable without sacrificing search quality.

    \item \textbf{\wb{} (right):} The evaluation suite benchmarks candidate workflows on three wireless task types. As shown in the right panel, the optimizer discovers distinct workflow topologies for various wireless tasks, such as the WCHW, WCNS and WCMSA workflows.
\end{enumerate}

\subsection{Domain-Specific Tool Library}
\label{sec:tools}

A key differentiator of \wap{} from general-purpose workflow optimizers is its \emph{domain-specific tool library} $\mathcal{F}$, which provides wireless-domain capabilities that LLMs inherently lack:

\subsubsection{Telecom Formula Retriever}
A retrieval-augmented generation (RAG) tool that searches a curated knowledge base of $31$ verified telecom formulas organized across 10+ categories: bandwidth (raised-cosine, NRZ, Nyquist), BER (BPSK, BFSK, ASK, DPSK, non-coherent BFSK, Nakagami-faded DPSK), FM/AM (Carson's rule, FM~SNR improvement factor $3\beta^2(\beta+1)$), delta modulation, PCM~(SQNR $= 6.02n + 1.76$~dB), Rayleigh fading (level crossing rate, average fade duration, Markov correlation $J_0(2\pi f_D T_s)$), channel capacity (Shannon), error-control coding (detection capability $d_{\min}-1$, correction capability $\lfloor(d_{\min}-1)/2\rfloor$), NOMA/SIC, and water-filling power allocation.

Given a natural-language query (e.g., ``BPSK BER over AWGN''), the retriever builds an inverted keyword index at initialization and scores each formula by a weighted relevance function:
\begin{equation}
    \mathrm{score}(q, f) = 2.0\,w_{\text{kw}} + 1.5\,w_{\text{name}} + 0.5\,w_{\text{notes}} + 0.3\,w_{\text{tex}} + 1.0\,w_{\text{cat}},
    \label{eq:rag_score}
\end{equation}
where $w_{\text{kw}}, w_{\text{name}}, w_{\text{notes}}, w_{\text{tex}} \in \{0,1\}$ indicate whether the query matches the formula's keywords, name, usage notes, or LaTeX body, and $w_{\text{cat}}$ is a category bonus for domain-aware synonym expansion (e.g., ``bit error'' $\to$ ``BER''). The top-$k$ formulas are returned with variable definitions, LaTeX representation, and notes.

\subsubsection{Telecom Calculator}
A SciPy-backed precision calculator that provides exact numerical computation for $20$ operations commonly encountered in wireless communications. The supported operations are: \texttt{erfc}, \texttt{erf}, $Q$-function, $Q^{-1}$-function, Bessel functions ($J_n$, $I_n$, $Y_n$), Marcum $Q$-function, BER computations (BPSK, BFSK, non-coherent BFSK, DPSK), Shannon capacity, Rayleigh fading statistics (outage probability, level crossing rate, average fade duration), Rician outage probability, and FM Bessel coefficients.

This tool addresses a critical weakness of LLMs, which frequently make numerical errors on transcendental functions. For the Marcum $Q$-function, the calculator first attempts \texttt{scipy.special.marcumq}; if unavailable, it falls back to numerical integration via \texttt{scipy.integrate.quad} with an upper limit of $\max(b+20, a+20, 50)$, or a convergent series expansion (up to 100 terms, convergence threshold $10^{-15}$). All computations use 10-digit precision.

\subsubsection{Ray-Tracing Channel Predictor}
\label{sec:ray_tracing_tool}
A site-specific ray-tracing engine estimates received signal quality based on user positions, enabling realistic channel-aware resource allocation. The engine operates on real-world building geometry extracted from OpenStreetMap (OSM) data covering three HKUST campus regions (North, Center, South). The key parameters are summarized in Table~\ref{tab:ray_tracing_params}.

\begin{table}[t]
\centering
\caption{Ray-tracing channel predictor parameters.}
\label{tab:ray_tracing_params}
\footnotesize
\begin{tabular}{@{}ll@{}}
\toprule
\textbf{Parameter} & \textbf{Value} \\
\midrule
TX power          & 30~dBm \\
Carrier frequency  & 2.4~GHz \\
Bandwidth          & 20~MHz \\
Noise figure       & 8~dB \\
Thermal noise      & $-174 + 10\log_{10}(\text{BW})$~dBm \\
User height        & 1.5~m \\
Default building height & 10.0~m (or OSM \texttt{levels} $\times 3.0$) \\
TX placement       & Centroid of tallest building $+$ 5.0~m \\
\bottomrule
\end{tabular}
\end{table}

Internally, the tool first determines the line-of-sight (LOS) condition by testing ray--building polygon intersections at fractional distances $t \in \{0.3, 0.5, 0.7\}$ between the transmitter and user. The path loss is then computed as:
\begin{equation}
    \text{PL}_{\text{LOS}} = 20\log_{10}(d) + 20\log_{10}(f) - 147.55 \;\text{(dB)},
    \label{eq:pl_los}
\end{equation}
\begin{equation}
    \text{PL}_{\text{NLOS}} = \text{PL}_{\text{LOS}} + 20 + 30\log_{10}\!\bigl(\max(d/100,\; 0.1)\bigr) \;\text{(dB)},
    \label{eq:pl_nlos}
\end{equation}
where $d$~(m) is the 3D distance and $f$~(Hz) is the carrier frequency. The received SNR~(dB) is obtained as $\text{SNR} = P_{\text{TX}} - \text{PL} - N_0$, and is then quantized to a CQI $\in \{1, \dots, 15\}$ via a linear mapping:
\begin{equation}
    \text{CQI} = \mathrm{round}\!\left(1 + 14 \cdot \frac{\mathrm{clamp}(\text{SNR},\, -10,\, 30) + 10}{40}\right).
    \label{eq:cqi_mapping}
\end{equation}
The full CQI-to-spectral-efficiency mapping follows the 4-bit CQI Table~1 of 3GPP TS~38.214 (Table~5.2.2.1-2) and is provided in Appendix~\ref{app:cqi_table}. Throughput is then $R = B \cdot \eta$, where $B$~(MHz) is the allocated bandwidth and $\eta$~(bps/Hz) is the spectral efficiency corresponding to the estimated CQI.

This tool supports both \texttt{ToolAgent} (LLM-driven) and \texttt{CodeLevel} (deterministic) invocation modes (cf.\ Remark~1); tool-assisted CQI prediction is critical, as direct LLM estimation yields near-random accuracy.

\subsubsection{Kalman Filter Predictor}
\label{sec:kalman_tool}
For the mobile service assurance benchmark (WCMSA), a code-level Kalman filter predicts future user positions from historical trajectory data. The filter uses a constant-velocity state-space model with state vector $\mathbf{x} = [x, y, v_x, v_y]^T$:
\begin{equation}
    \mathbf{x}_{t+1} = \mathbf{F}\,\mathbf{x}_t + \mathbf{w}_t, \quad
    \mathbf{z}_t = \mathbf{H}\,\mathbf{x}_t + \mathbf{v}_t,
    \label{eq:kalman_model}
\end{equation}
where the state transition and observation matrices are:
\begin{equation}
    \mathbf{F} = \begin{bmatrix} 1 & 0 & \Delta t & 0 \\ 0 & 1 & 0 & \Delta t \\ 0 & 0 & 1 & 0 \\ 0 & 0 & 0 & 1 \end{bmatrix}, \quad
    \mathbf{H} = \begin{bmatrix} 1 & 0 & 0 & 0 \\ 0 & 1 & 0 & 0 \end{bmatrix},
    \label{eq:kalman_FH}
\end{equation}
with $\Delta t = 1.0$~s. The process noise covariance is:
\begin{equation}
    \mathbf{Q} = q^2 \begin{bmatrix} \Delta t^4/4 & 0 & \Delta t^3/2 & 0 \\ 0 & \Delta t^4/4 & 0 & \Delta t^3/2 \\ \Delta t^3/2 & 0 & \Delta t^2 & 0 \\ 0 & \Delta t^3/2 & 0 & \Delta t^2 \end{bmatrix},
    \label{eq:kalman_Q}
\end{equation}
with $q = 0.5$, measurement noise $\mathbf{R} = r^2 \mathbf{I}_2$ with $r = 0.1$, and initial covariance $\mathbf{P}_0 = 10 \cdot \mathbf{I}_4$. The filter processes a trajectory of 4--6 historical positions to produce a one-step-ahead position prediction.

This tool is chained with the ray-tracing predictor within the \texttt{ToolAgent}'s ReAct loop: the agent first calls the Kalman filter to forecast the next position, then calls ray-tracing at the predicted coordinates to estimate future CQI, enabling \emph{proactive} resource management.

\subsection{MCTS-Based Workflow Optimization}
\label{sec:mcts}

Algorithm~\ref{alg:mcts} presents the overall optimization procedure:

\begin{algorithm}[t]
\caption{Refined MCTS for Workflow Optimization}
\label{alg:mcts}
\small
\begin{algorithmic}[1]
\Require Task $\mathcal{T}$, seed workflow $\mathcal{W}_0$, max rounds $T$, validation runs $V$
\Ensure Best workflow $\mathcal{W}^*$
\State Initialize: evaluate $\mathcal{W}_0$ on $\mathcal{D}_{\mathrm{val}}$, store result
\For{$t = 1, 2, \ldots, T$}
    \State Load processed experience $\mathcal{E}$
    \State $\mathcal{P} \gets$ \Call{GetTopRounds}{$K$} \Comment{Top-$K$ candidates}
    \State $\mathcal{W}_{\mathrm{parent}} \gets$ \Call{PenalizedBoltzmannSelect}{$\mathcal{P}$, $\mathcal{E}$}
    \State $\mathcal{R}_{\mathrm{critic}} \gets$ \Call{HeuristicCritic}{$\mathcal{W}_{\mathrm{parent}}$}
    \State $\hat{\mathcal{W}} \gets$ \Call{LLMMutate}{$\mathcal{W}_{\mathrm{parent}}$, $\mathcal{E}$, $\mathcal{R}_{\mathrm{critic}}$}
    \If{\Call{CheckModification}{$\hat{\mathcal{W}}$, $\mathcal{E}$} = \texttt{False}}
        \State \textbf{goto} Line 5 \Comment{Reject blocked mutations}
    \EndIf
    \State $\mathbf{s} \gets []$
    \For{$v = 1, \ldots, V$} \Comment{Multiple validation runs}
        \State $s_v \gets$ \Call{Evaluate}{$\hat{\mathcal{W}}$, $\mathcal{D}_{\mathrm{val}}$}
        \State Append $s_v$ to $\mathbf{s}$
    \EndFor
    \State $\bar{s} \gets \mathrm{median}(\mathbf{s})$ \Comment{Robust scoring}
    \State \Call{UpdateExperience}{$\mathcal{W}_{\mathrm{parent}}$, $\hat{\mathcal{W}}$, $\bar{s}$} \Comment{3-class}
    \State Store $(\hat{\mathcal{W}}, \bar{s})$
\EndFor
\State \Return $\mathcal{W}^* = \arg\max_t \bar{s}_t$
\end{algorithmic}
\end{algorithm}

\subsubsection{Selection (Penalized Boltzmann Sampling)}
\label{sec:selection}

Given the top-$K$ candidate workflows $\{(\mathcal{W}_i, s_i)\}_{i=1}^{K}$ sorted by average score $s_i$, we select a parent using a \emph{Penalized Boltzmann} distribution. Unlike the conventional MCTS method, our approach directly incorporates the exploration history through multiplicative penalty factors.

For each candidate $i$, we compute an exploration penalty $\rho_i \in [0.1, 1.3]$ based on its experience record:
\begin{equation}
    \rho_i = \max\!\Big(0.1,\; \min\!\big(1.3,\; 1 - 0.7 \cdot r_i^{\mathrm{fail}} + 0.2 \cdot r_i^{\mathrm{succ}} \big) \cdot \gamma_i \Big),
    \label{eq:penalty}
\end{equation}
where $r_i^{\mathrm{fail}} = n_i^{\mathrm{fail}}/n_i^{\mathrm{total}}$ is the failure ratio, $r_i^{\mathrm{succ}} = n_i^{\mathrm{succ}}/n_i^{\mathrm{total}}$ is the success ratio among all child mutations, and $\gamma_i = 0.8$ if $n_i^{\mathrm{total}} \geq 3$ (saturation discount), otherwise $\gamma_i = 1$.
The selection probability is then computed as a mixture of a uniform distribution and a penalized Boltzmann distribution:
\begin{equation}
    p_i = \lambda \cdot \frac{1}{K} + (1-\lambda) \cdot \frac{\rho_i \cdot \exp(\alpha \cdot \tilde{s}_i)}{\sum_{j=1}^{K} \rho_j \cdot \exp(\alpha \cdot \tilde{s}_j)},
    \label{eq:boltzmann}
\end{equation}
where $\tilde{s}_i = s_i - \max_j s_j$ is a shifted score for numerical stability, $\alpha > 0$ is a temperature parameter, and $\lambda \in (0,1)$ controls the exploration--exploitation trade-off.

\begin{remark}
The multiplicative penalty in~\eqref{eq:boltzmann} naturally implements ``soft pruning'': a node with all-failure history ($\rho_i \to 0.1$) is rarely selected but never completely removed, preserving the possibility of recovery. This is more robust than a hard prune, which is irreversible.
\end{remark}

\subsubsection{Expansion (LLM-Based Code Mutation)}
Given the selected parent workflow $\mathcal{W}_{\mathrm{parent}}$ with its code, prompt, score, error logs, and the critic report, an advanced Optimizer LLM generates a \emph{single focused modification}, \textit{e.g.}, revising a prompt, inserting a verification step, or changing the tool-use strategy. To maintain stability, each mutation is constrained to modify no more than 5 lines of code. The LLM is provided with:
i) The current workflow code and prompt; ii) The formatted experience (past successes, failures, neutral results); iii) The critic report with score-aware recommendations; iv) A description of available operators and tools. Before the mutation is accepted, a \emph{modification check} verifies that:
1)~it is not an exact duplicate of a previously failed modification,
2)~it does not contain known harmful patterns (\textit{e.g.,} patterns experimentally proven to degrade performance).

\subsubsection{Executing Evaluation}
Each candidate workflow is evaluated $V$ times on the validation set $\mathcal{D}_{\mathrm{val}}$ by a cost-efficient Executor LLM, and the \emph{median} score (rather than the mean) is used as the performance estimate. This provides robustness against outlier evaluations, which are common in wireless settings due to the stochastic nature of LLM outputs and the sensitivity of numerical computations. The evaluation records, including per-problem scores, error logs, and API costs, are stored for subsequent analysis.

\subsubsection{Experience Backpropagation}
The evaluation outcome is classified via the 3-class mechanism (Section~\ref{sec:experience}) and recorded alongside the modification description, failure logs, and scores. These accumulated experiences directly inform the penalty factor $\rho$ in future Boltzmann selection~\eqref{eq:penalty} and provide structured exemplars for the Optimizer LLM's next expansion.

\subsection{Maturity-Aware Heuristic Critic}
\label{sec:critic}

A straightforward approach would evaluate every proposed mutation on the full validation set. However, each evaluation requires multiple LLM API calls and is both time-consuming and expensive. To avoid unnecessary evaluations, we introduce a \emph{maturity-aware heuristic critic} that pre-screens mutations based on the current workflow's characteristics.

More importantly, the critic is \textit{not} an LLM; it is a lightweight, zero-cost, rule-based analyzer that examines:

\begin{enumerate}[leftmargin=*]
    \item \textbf{Performance maturity:} Based on the current score $s$, the critic determines the recommended aggressiveness level. Table~\ref{tab:critic_levels} summarizes the four-level policy, where the thresholds are set at $\tau_{\mathrm{high}} = 0.65$ and $\tau_{\mathrm{mid}} = 0.50$:
    \begin{equation}
        \mathrm{Level}(s) = \begin{cases}
            \text{Ultra-conservative}, & \text{node saturated}, \\
            \text{Conservative}, & s \geq \tau_{\mathrm{high}} = 0.65, \\
            \text{Moderate}, & \tau_{\mathrm{mid}} \leq s < \tau_{\mathrm{high}}, \\
            \text{Aggressive}, & s < \tau_{\mathrm{mid}} = 0.50.
        \end{cases}
        \label{eq:aggressiveness}
    \end{equation}

    \begin{table}[t]
    \centering
    \caption{Heuristic critic aggressiveness levels and allowed mutation types. A node is \emph{saturated} if it has $\geq 2$ failed mutations and zero successes.}
    \label{tab:critic_levels}
    \footnotesize
    \begin{tabular}{@{}lcp{4.2cm}@{}}
    \toprule
    \textbf{Level} & \textbf{Condition} & \textbf{Allowed Changes} \\
    \midrule
    Ultra-conservative & Saturated & Single word / punctuation only \\
    Conservative      & $s \geq 0.65$ & Minor prompt wording; no structural changes \\
    Moderate          & $0.50 \leq s < 0.65$ & Single structural change; no new branches \\
    Aggressive        & $s < 0.50$ & Structural changes, new operators, ToolAgent additions \\
    \bottomrule
    \end{tabular}
    \end{table}

    High-scoring workflows should receive only minor prompt adjustments; low-scoring workflows may undergo structural changes.

    \item \textbf{Structural complexity:} By parsing the workflow code via abstract syntax tree (AST) analysis, the critic detects the number of operator calls, presence of conditional logic, and tool usage patterns. Over-engineered workflows (e.g., $5+$ sequential steps, nested conditionals) are flagged, and the report recommends simplification.

    \item \textbf{Error pattern analysis:} By analyzing the evaluation log, the critic categorizes errors into format errors (answer extraction failures), unit errors (wrong unit conversion), and value errors (incorrect computation), then provides targeted recommendations.

    \item \textbf{Node saturation:} If a workflow has been mutated $\geq 2$ times with all failures and no successes, it is marked as \emph{saturated}. The aggressiveness is overridden to \texttt{ULTRA-CONSERVATIVE}, and the optimizer is advised to select a different parent.
\end{enumerate}

The critic report is appended to the LLM mutation prompt, guiding, but not constraining, the Optimizer LLM's proposals.

\subsection{3-Class Experience Replay}
\label{sec:experience}

In wireless benchmarks, evaluation scores are subject to noise from LLM stochasticity, numerical precision, and answer extraction variability. Standard two-class experience (success/failure) can misclassify marginal improvements as successes and marginal degradations as failures, leading the optimizer to ``chase noise.'' Thus, we introduce a significance threshold $\epsilon > 0$ and classify each mutation outcome into three classes:
\begin{equation}
    \mathrm{Class}(\Delta s) = \begin{cases}
        \textsc{Success}, & \Delta s > +\epsilon, \\
        \textsc{Neutral}, & |\Delta s| \leq \epsilon, \\
        \textsc{Failure}, & \Delta s < -\epsilon,
    \end{cases}
    \label{eq:3class}
\end{equation}
where $\Delta s = \bar{s}_{\mathrm{new}} - s_{\mathrm{parent}}$ is the score change. In our experiments, we set $\epsilon = 0.02$ (i.e., 2\% absolute accuracy).

These classes are treated differently in experience replay:
\begin{itemize}[leftmargin=*]
    \item \textbf{Success:} The modification is stored as a positive exemplar. The optimizer is encouraged to build upon similar strategies.
    \item \textbf{Neutral:} The modification is recorded but not penalized. Minor variations of neutral approaches are allowed, acknowledging that the direction may be promising with different parameters.
    \item \textbf{Failure:} The exact modification is blacklisted to prevent repetition. The failure also contributes to the parent node's penalty factor $\rho$ in~\eqref{eq:penalty}.
\end{itemize}

\begin{remark}
The choice of $\epsilon$ balances sensitivity and stability. Too small an $\epsilon$ makes the optimizer reactive to noise; too large an $\epsilon$ causes it to miss genuine improvements. We analyze the sensitivity to $\epsilon$ in Section~\ref{sec:ablation}.
\end{remark}

\subsection{Convergence Detection}
\label{sec:convergence}
To avoid unnecessary computation, \wap{} monitors convergence by tracking the running top-$k$ average score across rounds. Specifically, let $\bar{S}_t$ denote the mean of the top-$k$ scores observed up to round $t$, and let $\sigma_{\bar{S}_t} = \sqrt{\sum_{j \in \text{top-}k} \sigma_j^2 / k^2}$ be the associated standard error computed from the per-round score variances (obtained from the $V$ validation runs). We define the inter-round improvement and its uncertainty as:
\begin{equation}
    \Delta_t = \bar{S}_t - \bar{S}_{t-1}, \quad \sigma_{\Delta_t} = \sqrt{\sigma_{\bar{S}_t}^2 + \sigma_{\bar{S}_{t-1}}^2}.
    \label{eq:convergence}
\end{equation}
The search terminates early if $|\Delta_t| \leq z \cdot \sigma_{\Delta_t}$ for $C$ consecutive rounds, where $z$ is a significance level (default $z = 0$, i.e., strictly no improvement) and $C = 5$ is the patience parameter. In our experiments, we set $k = 3$. This criterion ensures that the optimizer stops only when the top-performing workflows have genuinely plateaued, rather than reacting to single-round fluctuations.

\section{\wb{}: A Comprehensive Wireless Benchmark}
\label{sec:benchmark}

Existing LLM benchmarks (e.g., MMLU, GSM8K, HumanEval) do not capture the unique challenges of wireless communication tasks: domain-specific formulas, unit-sensitive numerical reasoning, tool-assisted computation, and multi-step network management decisions. We introduce \wb{}\footnote{The code is
available at https://github.com/jwentong/WirelessBench.} to fill this gap.

\subsection{Design Principles}
\wb{} covers three complementary dimensions of wireless AI capability:
\begin{itemize}[leftmargin=*]
    \item \textbf{Knowledge Reasoning} (WCHW): Can the agent \emph{understand and apply} wireless communication theory?
    \item \textbf{Intent Understanding} (WCNS): Can the agent \emph{understand correct user intent} that interact with wireless environment?
    \item \textbf{Multi-Step Decision Making} (WCMSA): Can the agent \emph{predict, estimate, and allocate} resources for mobile users with proactive service assurance?
\end{itemize}

\subsection{Data Construction Pipeline}
\label{sec:data_pipeline}
All three \wb{} benchmarks are constructed through a unified four-stage pipeline, as summarized below.

\begin{enumerate}[leftmargin=*]
    \item \textbf{Data collection.} Seed problems are curated from authoritative sources: wireless communication textbooks~\cite{goldsmith2005wireless, molisch2012wireless} for WCHW, and 3GPP/IEEE standards (e.g., 3GPP TS~38.214, TS~38.901) for WCNS and WCMSA. Each problem is paired with a step-by-step CoT solution.
    \item \textbf{Psychometric data cleaning.} Inspired by~\cite{truong2025fantastic}, we treat multiple LLMs as ``examinees'' and apply a funnel-style cleaning pipeline: (i)~rule-based pre-cleaning removes non-computational or incomplete items; (ii)~ten LLMs independently solve each problem, and a hierarchical five-level judge determines correctness; (iii)~three psychometric metrics, item-total correlation, Mokken scale analysis, and inter-item consistency, identify low-quality items; and (iv)~an advanced LLM auditor diagnoses flagged items, correcting erroneous samples and ambiguous ones.
    \item \textbf{LLM-based augmentation.} Multiple LLMs expand the cleaned corpus through parameter variation, bidirectional conversion, cross-topic integration, and boundary exploration, followed by TF-IDF-based deduplication. Importantly, LLMs are used \emph{only} to generate new problem text and parameter variations; all ground-truth answers and chain-of-thought solutions are recomputed by deterministic numerical solvers or analytical formulas and subsequently verified by human annotators (Stage~4).
    \item \textbf{Human validation.} Graduate students verify every problem: checking solution correctness, correcting chain-of-thought reasoning, and standardizing answer formats.
\end{enumerate}

\subsection{WCHW: Wireless Communication Homework}
WCHW consists of 1,392 problems (348 validation, 1,044 test) drawn from university-level wireless communication textbooks, covering topics such as modulation (BPSK, QPSK, QAM), channel capacity (Shannon, ergodic), error probability (coherent/non-coherent detection), signal processing (PCM, delta modulation), and antenna theory (see Appendix~\ref{Ex_WCHW} for representative examples). Each problem requires multi-step reasoning and involves:
(1) Application of domain-specific formulas (e.g., $P_e = \frac{1}{2}\mathrm{erfc}\!\left(\!\sqrt{E_b/N_0}\right)$ for BPSK);
(2) Unit conversion (e.g., kHz $\to$ Hz, dBm $\to$ W);
(3) Numerical computation with special functions.

\textbf{Answer format classification.}
Each reference answer is first classified into one of nine format types: \textsc{PureNumeric}, \textsc{NumericWithUnit}, \textsc{Scientific}, \textsc{Formula}, \textsc{Percentage}, \textsc{TextShort}, \textsc{TextLong}, \textsc{CodeSequence}, and \textsc{Ratio}. A \texttt{QuestionAnalyzer} uses regex-based pattern matching on the question text to predict the expected answer format, including unit detection (frequency, data rate, time, power, distance, spectral efficiency) and fill-in-the-blank detection.

\textbf{Multi-strategy scoring.}
The evaluation pipeline applies four scoring strategies in parallel and returns the \emph{maximum} score, reducing false negatives from format mismatches. Table~\ref{tab:wchw_scoring} summarizes the numeric scoring tiers; formula answers are compared via LaTeX normalization and character-level Jaccard similarity ($>0.7 \to 0.8$, $>0.5 \to 0.5$); text answers are scored by weighted keyword overlap (60\% number match + 40\% word match). The system automatically recognizes 38 unit multipliers across six physical dimensions (frequency, data rate, power, time, distance, spectral efficiency) and converts all values to base SI units before comparison.

\begin{table}[t]
\centering
\caption{WCHW numeric scoring tiers. The system also detects common errors such as factor-of-$10^3$ and factor-of-$2$ mistakes.}
\label{tab:wchw_scoring}
\footnotesize
\begin{tabular}{@{}lc@{}}
\toprule
\textbf{Condition} & \textbf{Score} \\
\midrule
Relative error $< 1\%$ & 1.0 \\
Relative error $< 5\%$ & 0.9 \\
Relative error $< 10\%$ & 0.7 \\
Off by $\times 10^3$ or $\times 10^6$ (unit error) & 0.5 \\
Off by $\times 2$ (common miscalculation) & 0.3 \\
Otherwise & 0.0 \\
\bottomrule
\end{tabular}
\end{table}

\subsection{WCNS: Wireless Communication Network Slicing}
WCNS models a 5G network slicing scenario with two slice types: eMBB (enhanced Mobile Broadband, 90~MHz capacity) and URLLC (Ultra-Reliable Low-Latency Communication, 30~MHz capacity). Each problem presents (see Appendix~\ref{Ex_WCNS} for an example):
(1) A network state (number of active users per slice),
(2) A new user's position (2D coordinates on the HKUST campus) and a natural-language service request, and
(3) The requirement to \emph{predict the user's CQI} using a ray-tracing tool before making a slicing decision.

\textbf{Dataset generation.}
User positions are sampled uniformly at random from outdoor regions (verified outside building footprints via a ray-casting point-in-polygon test) across three HKUST campus regions using real OpenStreetMap building data. The number of existing users is drawn as $n_{\text{eMBB}} \sim \mathcal{U}[2,12]$ and $n_{\text{URLLC}} \sim \mathcal{U}[2,10]$. For each problem, a service request is randomly selected from a bank of $15$ eMBB and $15$ URLLC natural-language templates (e.g., ``I need to stream a 4K video'' $\to$ eMBB; ``autonomous vehicle V2X communication'' $\to$ URLLC). The ground-truth CQI is computed by the ray-tracing engine (Section~\ref{sec:ray_tracing_tool}), and bandwidth is allocated via proportional fairness: $B = B_{\text{total}} / (n + 1)$, clamped to $[B_{\min}, B_{\max}]$ per slice type.

\textbf{Agent pipeline.}
For the task, an agent must: (1)~classify the service type (eMBB vs.\ URLLC), (2)~invoke the ray-tracing predictor \emph{via the ReAct-based ToolAgent} to obtain accurate CQI values, (3)~calculate resource allocation following the proportional-fairness policy, and (4)~output a structured decision.

\textbf{Composite scoring.}
Table~\ref{tab:wcns_scoring} defines the 4-metric weighted scoring function. Bandwidth and throughput use a tiered relative-error scheme (see the rightmost column); CQI allows partial credit for near-misses.

\begin{table}[t]
\centering
\caption{WCNS composite scoring: 4-metric weighted scheme.}
\label{tab:wcns_scoring}
\footnotesize
\begin{tabular}{@{}lcl@{}}
\toprule
\textbf{Sub-Metric} & \textbf{Weight} & \textbf{Scoring Rule} \\
\midrule
Slice type & 25\% & Binary: exact match $\to$ 1.0, else 0.0 \\
CQI        & 15\% & Exact $\to$ 1.0; $\pm 1 \to$ 0.8; $\ldots$; $\pm 3{\text{--}}4 \to$ 0.2 \\
Bandwidth  & 35\% & Rel.\ err.:\ ${<}2\%{\to}1.0$;\ $\ldots$;\ ${<}20\%{\to}.70$ \\
Throughput & 25\% & Same tiered scheme as bandwidth \\
\bottomrule
\end{tabular}
\end{table}

\subsection{WCMSA: Wireless Communication Mobile Service Assurance}
WCMSA extends the WCNS scenario to include \emph{user mobility}, requiring the agent to predict future positions, estimate channel conditions at predicted locations, and make proactive resource management decisions (see Appendix~\ref{Ex_WCMSA} for an example).

\textbf{Trajectory generation.}
Each mobile user follows a constant-velocity trajectory with Gaussian perturbations. A start position is sampled uniformly in an outdoor region at distance $50$--$150$~m from the base station; the walking speed is drawn from $\mathcal{U}[1.0, 2.5]$~m/s, and position noise $\sim \mathcal{N}(0, 0.15)$~m is added at each step. The direction undergoes small angular noise $\sim \mathcal{N}(0, 0.1)$~rad with a $10\%$ probability of a large turn ($\pm[0.3, 0.8]$~rad). Each problem provides $4$--$6$ historical positions (including the current position) and requires prediction of the next position.

\textbf{Service types and QoS requirements.}
WCMSA defines $10$ eMBB and $10$ URLLC service types, each with a minimum throughput requirement. Representative services include ``4K video streaming'' (eMBB, $\geq$25~Mbps), ``cloud gaming with high graphics'' (eMBB, $\geq$35~Mbps), ``8K live sports broadcast'' (eMBB, $\geq$80~Mbps), ``remote robotic surgery'' (URLLC, $\geq$10~Mbps), ``autonomous vehicle V2X'' (URLLC, $\geq$5~Mbps), and ``real-time drone control'' (URLLC, $\geq$8~Mbps). The full service type list and QoS requirements are provided in Appendix~\ref{app:service_types}.

\textbf{Agent pipeline.}
Each problem provides a time series of user positions, current network state (active users per slice), a service request, and a minimum QoS requirement. The agent must: (1)~predict the user's next position via the Kalman filter tool, (2)~estimate the CQI at the \emph{predicted} (not current) position via the ray-tracing tool---both invoked through the ReAct-based \texttt{ToolAgent} in a chained reasoning-acting loop, (3)~classify the service type and allocate bandwidth, (4)~compute expected throughput, and (5)~determine whether QoS requirements will be satisfied.

\textbf{Composite scoring.}
Table~\ref{tab:wcmsa_scoring} defines the 6-metric weighted scoring function. Position prediction uses a smooth distance-decay function $\mathrm{score} = \max(0, 1 - (d/20)^{1.2})$, where $d$ is the Euclidean distance between predicted and ground-truth positions; score reaches zero at $d \geq 20$~m. QoS verification (5\% weight) is treated as a bonus metric.

\begin{table}[t]
\centering
\caption{WCMSA composite scoring: 6-metric weighted scheme.}
\label{tab:wcmsa_scoring}
\footnotesize
\begin{tabular}{@{}lcl@{}}
\toprule
\textbf{Sub-Metric} & \textbf{Weight} & \textbf{Scoring Rule} \\
\midrule
Position prediction  & 15\% & Distance decay: $\max(0, 1 - (d/20)^{1.2})$ \\
CQI prediction       & 15\% & Exact $\to$ 1.0; $\ldots$; $\pm 3 \to$ 0.50 \\
Slice type           & 20\% & Binary: exact $\to$ 1.0, else 0.0 \\
Bandwidth allocation & 25\% & Same tiered scheme as WCNS \\
Throughput           & 20\% & Same tiered scheme as WCNS \\
QoS verification     & 5\%  & Binary: correct Yes/No $\to$ 1.0, else 0.0 \\
\bottomrule
\end{tabular}
\end{table}

\subsection{Dataset Statistics and Oracle Construction}
Table~\ref{tab:benchmark_stats} summarizes the \wb{} datasets. All ground truths are constructed from deterministic rules or expert solutions, ensuring reproducibility.

\begin{table}[t]
\centering
\caption{Statistics of the \wb{} benchmark suite.}
\label{tab:benchmark_stats}
\footnotesize
\begin{tabular}{@{}lcccl@{}}
\toprule
\textbf{Benchmark} & \textbf{Val.} & \textbf{Test} & \textbf{Metric} & \textbf{Task Type} \\
\midrule
WCHW & 348 & 1,044 & Accuracy & Knowledge reasoning \\
WCNS & 250 & 750 & Composite & Code + tool use \\
WCMSA & 250 & 750 & Composite & Multi-step diagnosis \\
\bottomrule
\end{tabular}
\end{table}

\section{Experimental Results}
\label{sec:experiments}

\subsection{Experimental Setup}
\label{sec:setup}

\subsubsection{LLM Backbones}
We use Claude-Opus-4.5 as the Optimizer LLM (for generating workflow mutations) and Qwen-turbo-latest as the Executor LLM (for running the actual workflows). This separation facilitates lightweight deployment in practice and allows us to study cost-performance trade-offs.

\subsubsection{Baselines}
We compare \wap{} against the following baselines on both general NLP benchmarks (HotpotQA, DROP, MATH) and the \wb{}:
\begin{itemize}[leftmargin=*]
    \item \textbf{Qwen-turbo-latest~\cite{qwen2024qwen25}:} Zero-shot prompting with the Qwen-turbo model, serving as a non-agentic LLM baseline.
    \item \textbf{CoT~\cite{wei2022cot}:} Chain-of-thought prompting that elicits step-by-step reasoning before producing the final answer.
    \item \textbf{MedPrompt~\cite{nori2023medprompt}:} A state-of-the-art prompting strategy combining CoT, self-consistency, and ensemble selection.
    \item \textbf{ADAS~\cite{hu2024adas}:} Automated Design of Agentic Systems, a concurrent agent architecture search method.
    \item \textbf{AFlow~\cite{zhang2025aflow}:} MCTS-based workflow optimization without our domain-specific enhancements (ToolAgent, Penalized Boltzmann Selection, Heuristic Critic, 3-Class Experience).
\end{itemize}
For AFlow and \wap{}, the optimization model is Claude-Opus-4.5 and the execution model is Qwen-turbo-latest. In addition, for the WCNS and WCMSA case studies, we compare against \textbf{WirelessAgent}~\cite{tong2024wirelessagent}, our prior hand-designed, fixed-workflow baseline that uses the same domain tools (ray-tracing predictor, Kalman filter, precision calculator) within a manually engineered ReAct pipeline.

\subsubsection{Hyperparameters}
All experiments use the following default settings unless otherwise stated: top-$K = 5$ candidates for selection, $\lambda = 0.3$ for exploration mixing, $\alpha = 0.2$ for Boltzmann temperature, $\epsilon = 0.02$ for 3-class threshold, $V = 5$ validation runs per evaluation, and a maximum of $T = 20$ optimization rounds. All LLM models are accessed via APIs, and the temperature is set to $0$ for all models.

\subsection{Main Results}
\label{sec:main_results}
Table~\ref{tab:main_results} compares \wap{} with baselines on three general NLP benchmarks (HotpotQA, DROP, MATH) and the wireless-domain benchmark \wb{}.  \wap{} achieves the highest scores on every benchmark, reaching F1 scores of $0.7273$ on HotpotQA and $0.8021$ on DROP, and solve rates of $0.8210$ on MATH and $0.8102$ on \wb{}~(After). These consistent gains across both general and domain-specific tasks confirm that the MCTS-based workflow optimizer generalizes well.
In addition, the improvements over AFlow, the strongest automated baseline, are particularly pronounced on the wireless benchmark: \wap{} outperforms AFlow by $20.12$~pp on \wb{}~(Before) and $11.10$~pp on \wb{}~(After). This gap highlights the value of our three domain-aware enhancements and \texttt{ToolAgent} which together enable the optimizer to navigate the noisy, tool-dependent evaluation landscape of wireless tasks far more effectively.
Finally, the ``After'' column, which reflects the psychometric data-cleaning pipeline (Section~\ref{sec:data_pipeline}), yields higher scores than ``Before'' with \wap{} benefiting from an additional $9.38$~pp boost. This underscores that high-quality benchmark data and robust evaluation are essential complements to workflow optimization.

\textbf{Search cost.} The entire MCTS optimization is remarkably inexpensive: the total API cost for the full search is \$4.95 on WCHW (19~rounds, $\sim$63~min wall-clock), \$0.99 on WCNS (11~rounds, $\sim$13~min), and \$1.05 on WCMSA (11~rounds, $\sim$14~min). At inference time with the optimized workflow, the per-problem cost is $<$\$0.001 for all three benchmarks (\$0.00083/problem on WCHW, \$0.00056/problem on WCNS, \$0.00068/problem on WCMSA), confirming that \wap{} is practical for deployment.

\begin{table*}[t]
\centering
\caption{Performance comparison on general NLP benchmarks and the wireless benchmark. \wb{}  ``Before'' uses the original dataset; ``After'' applies the psychometric data-cleaning pipeline (Section~\ref{sec:data_pipeline}) that detects and corrects low-quality samples. Best results in \textbf{bold}.}
\label{tab:main_results}
\footnotesize
\begin{tabular}{@{}l ccc cc@{}}
\toprule
 & & & & \multicolumn{2}{c}{\textbf{\wb{} (Solve Rate)}} \\
\cmidrule(lr){5-6}
\textbf{Method} & \textbf{HotpotQA (F1)} & \textbf{DROP (F1)} & \textbf{MATH (Solve Rate)} & Before & After \\
\midrule
Qwen-turbo-latest~\cite{qwen2024qwen25} & 0.1423 & 0.5641 & 0.6412 & 0.5013 & 0.5834 \\
CoT~\cite{wei2022cot} & 0.5379 & 0.7607 & 0.6828 & 0.5032 & 0.6032 \\
MedPrompt~\cite{nori2023medprompt} & 0.5411 & 0.7559 & 0.6996 & 0.5422 & 0.6122 \\
ADAS~\cite{hu2024adas} & 0.5110 & 0.7423 & 0.4953 & 0.4813 & 0.5313 \\
AFlow~\cite{zhang2025aflow} & 0.5823 & 0.7811 & 0.7864 & 0.5152 & 0.6992 \\
\midrule
\wap{} (Ours) & \textbf{0.7273} & \textbf{0.8021} & \textbf{0.8210} & \textbf{0.7164} & \textbf{0.8102} \\
\bottomrule
\end{tabular}
\end{table*}

\subsection{Case Studies}
\label{sec:case_studies}

We present detailed case studies on the three \wb{} benchmarks, examining both quantitative performance and the qualitative evolution of workflows discovered by the MCTS optimizer.

\subsubsection{WCHW (Knowledge Reasoning with Tool Verification)}

\begin{figure}[t]
    \centering
    \includegraphics[width=\columnwidth]{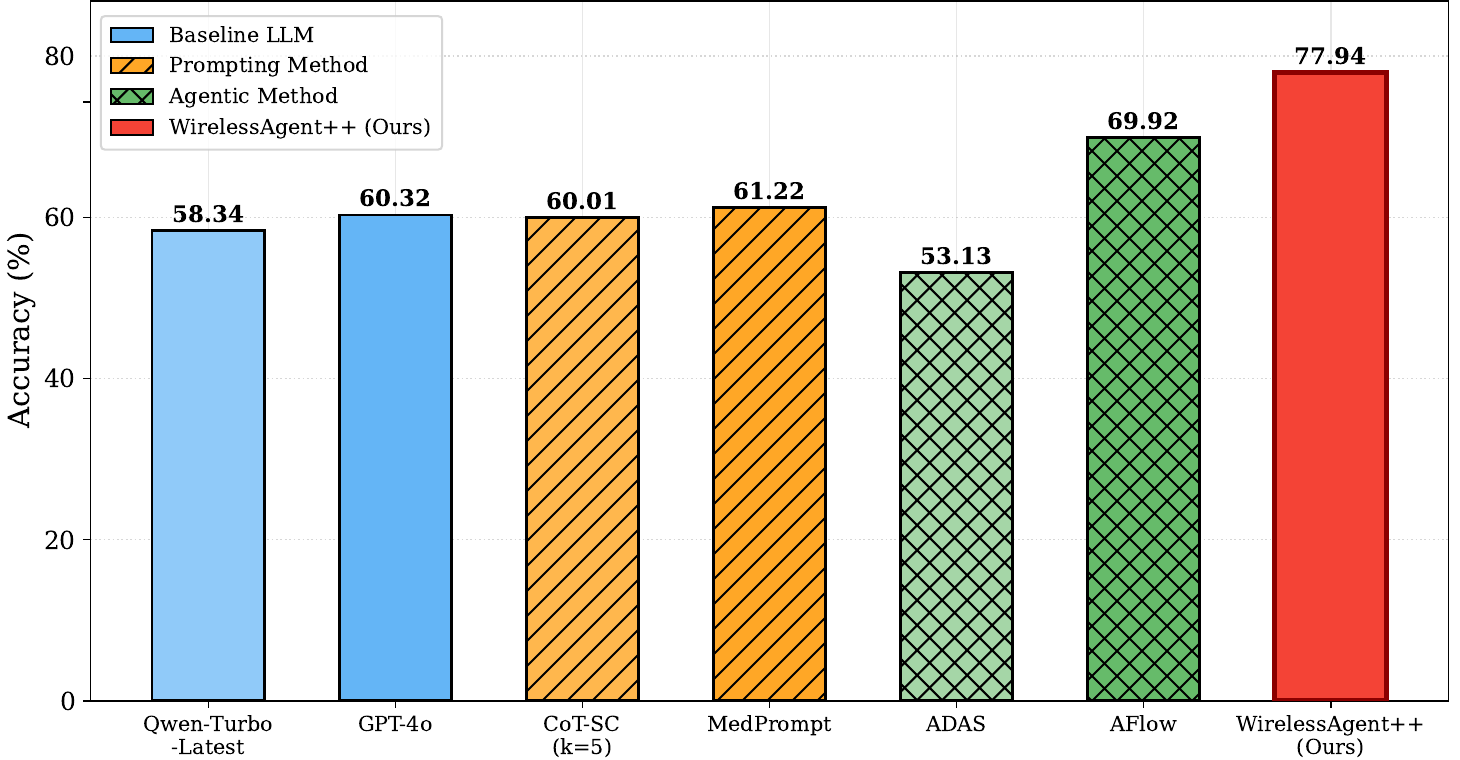}
    \caption{Method comparison on the test dataset of the WCHW benchmark. \wap{} outperforms the AFlow and all prompting baselines.}
    \label{fig:wchw_comparison}
\end{figure}

Fig.~\ref{fig:wchw_comparison} compares different methods on the WCHW benchmark. \wap{} significantly outperforms all baselines, demonstrating the effectiveness of domain-specific agentic workflow optimization for wireless knowledge reasoning.

\begin{figure}[t]
    \centering
    \includegraphics[width=\columnwidth]{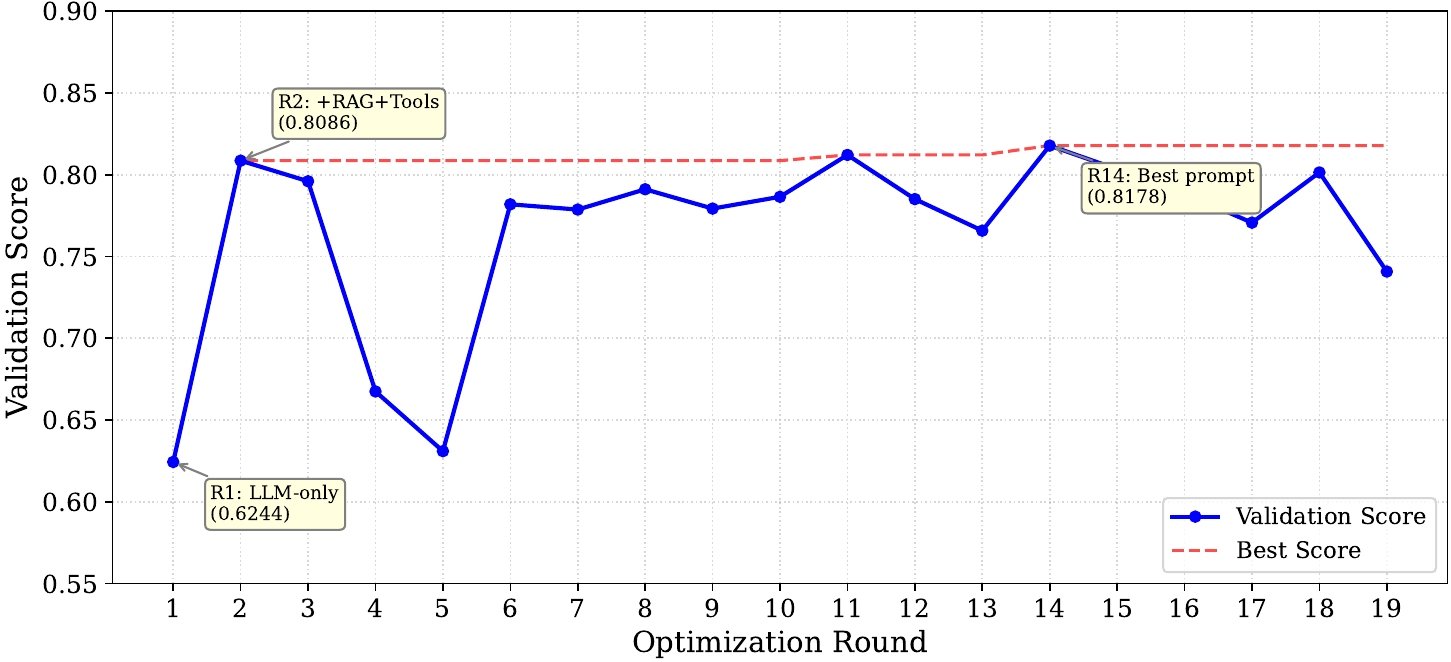}
    \caption{Workflow evolution of \wap{} on the validation dataset of the WCHW benchmark.}
    \label{fig:wchw_we}
\end{figure}
\textbf{Workflow evolution.} Fig.~\ref{fig:wchw_we} shows the workflow evolution of \wap{} on the validation dataset of the WCHW benchmark, where the optimizer follows a three-phase trajectory.
\emph{Phase~1---Seed (Round~1, 62.44\%):} A single \texttt{Custom} operator with a manually curated formula library covering ${\sim}10$ telecom topics (DM~SNR, PCM~SQNR, coherent/non-coherent BER, FM modulation, Shannon capacity, Rayleigh fading). Despite the rich prompt, accuracy is limited by LLM errors on transcendental-function arithmetic.
\emph{Phase~2---Tool Discovery (Round~2, 80.86\%, $+$18.42~pp):} The optimizer introduces a \texttt{ToolAgent} verification step: (1)~\texttt{Custom} solves the problem using the formula library, then (2)~\texttt{ToolAgent} executes Python code to independently verify numerics and convert answers to base SI units. This \emph{single structural change---discovered automatically}---yields the largest improvement across all 19~rounds.
\emph{Phase~3---Prompt Refinement (Round~14, Best, 81.78\%):} The prompt is iteratively expanded with formulas for water-filling power allocation, matched-filter design, coherent-OOK optimal thresholds, amplifier IM3 characterization, and NOMA/SIC uplink capacity. The \texttt{ToolAgent} prompt is also refined to handle formula-type answers (e.g., $h(t) = A[U(t{-}2T/3) - U(t{-}T)]$), not just numerical outputs. The complete optimal workflow is detailed in Appendix~\ref{OW_WCHW}.

\subsubsection{WCNS (Ray-Tracing-Augmented Network Slicing)}

\begin{figure*}[!t]
\centering
\subfloat[Overall composite score comparison]{\label{fig:wcns_overall}
\includegraphics[width=0.55\columnwidth]{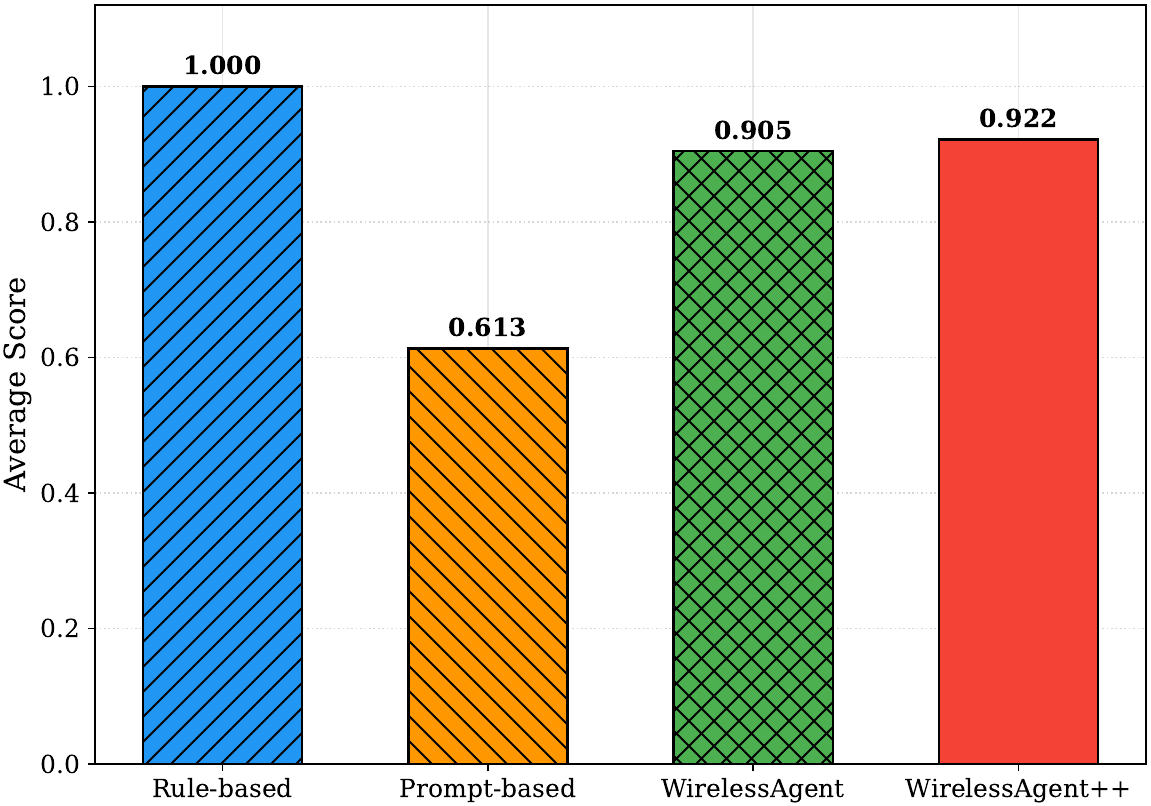}}
\hfil
\subfloat[Per-metric breakdown]{\label{fig:wcns_metric}
\includegraphics[width=0.83\columnwidth]{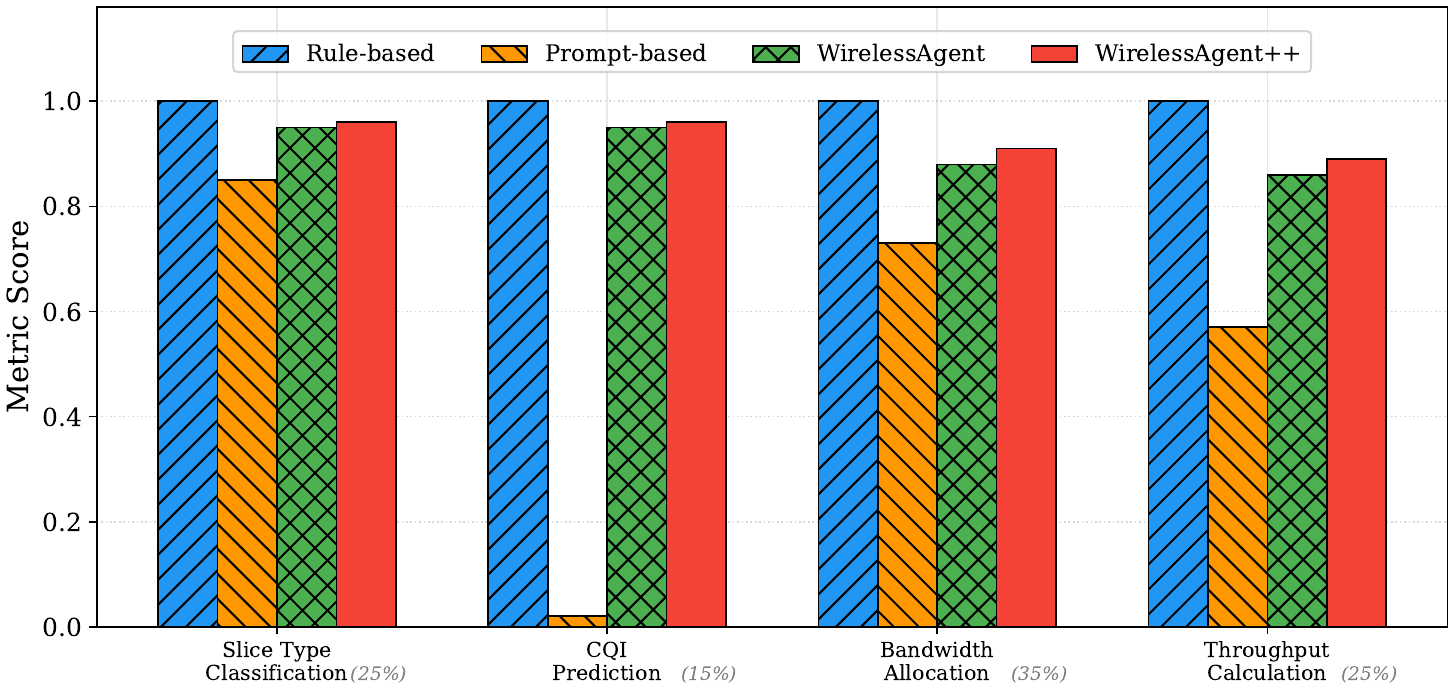}}
\hfil
\subfloat[Workflow evolution]{\label{fig:wcns_we}
\includegraphics[width=0.62\columnwidth]{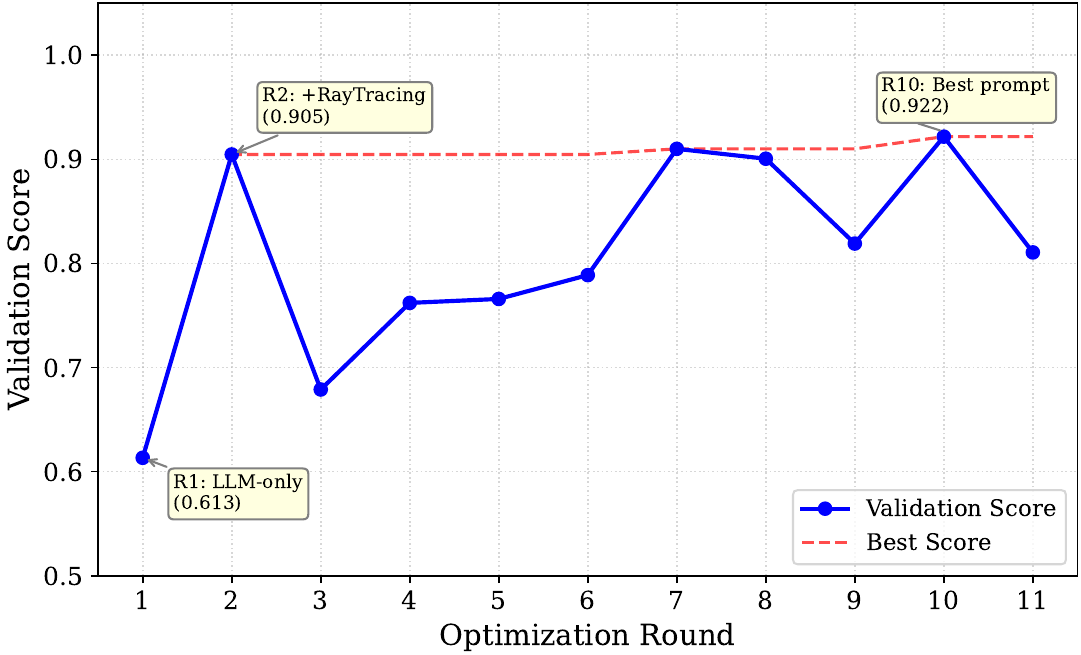}}
\caption{WCNS case study. (a) Overall composite score comparison across methods. (b) Per-metric breakdown: \wap{} achieves $\geq$89\% on all four sub-metrics. (c) Workflow evolution on the validation set showing the three-phase optimization trajectory.}
\label{fig:wcns}
\end{figure*}

\begin{figure*}[!t]
\centering
\subfloat[North scenario] {\label{fig:tp_north}
\includegraphics[width=0.65 \columnwidth]{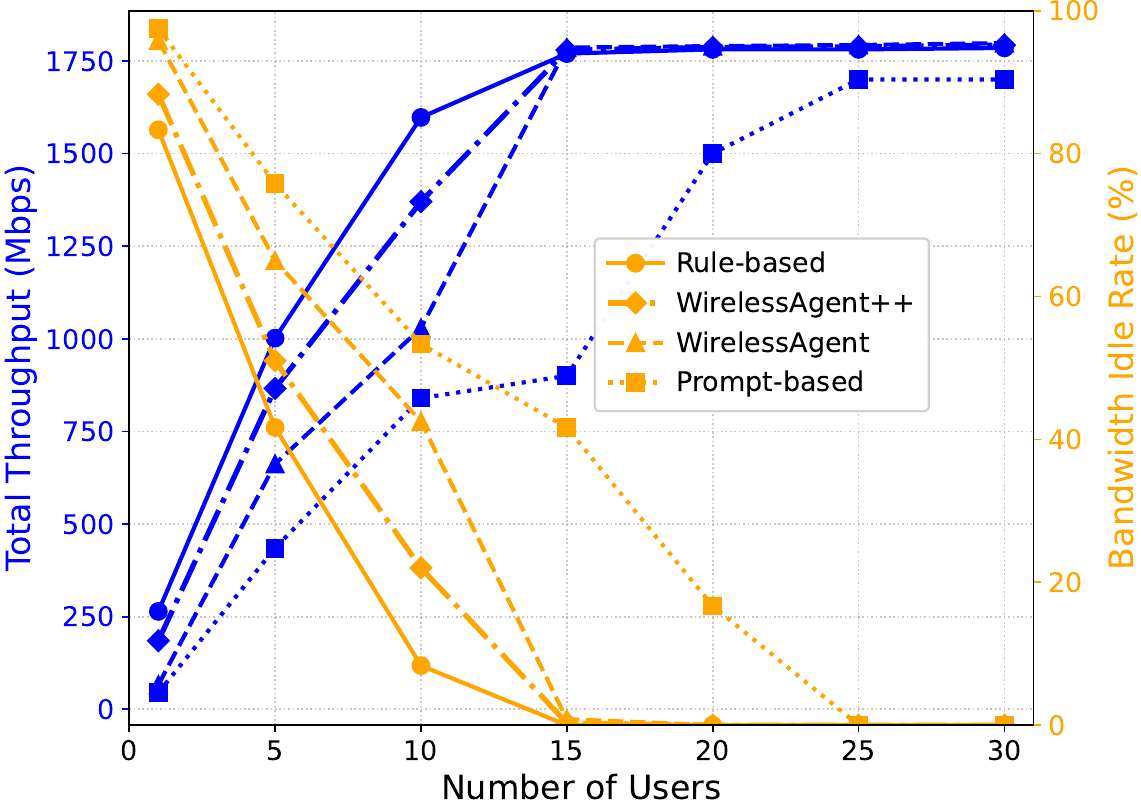}}
\hfil
\subfloat[Center scenario] {\label{fig:tp_center}
\includegraphics[width=0.65 \columnwidth]{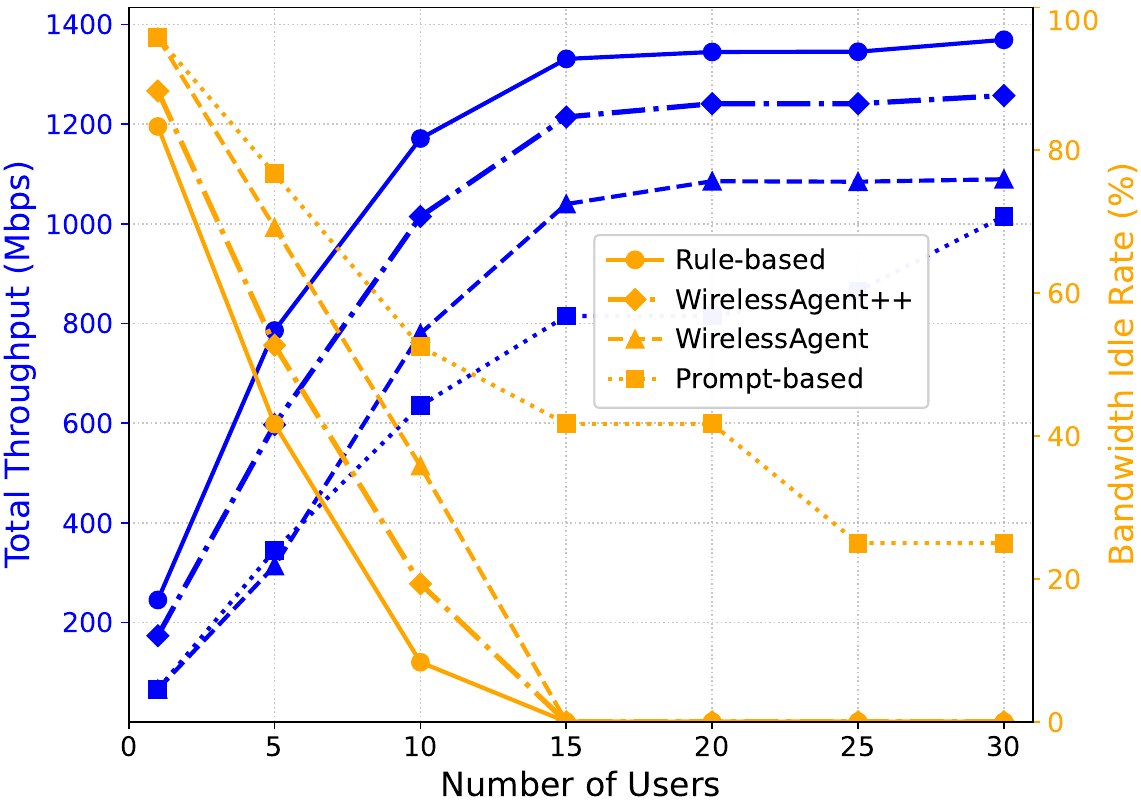}}
\hfil
\subfloat[South scenario] {\label{fig:tp_south}
\includegraphics[width=0.65 \columnwidth]{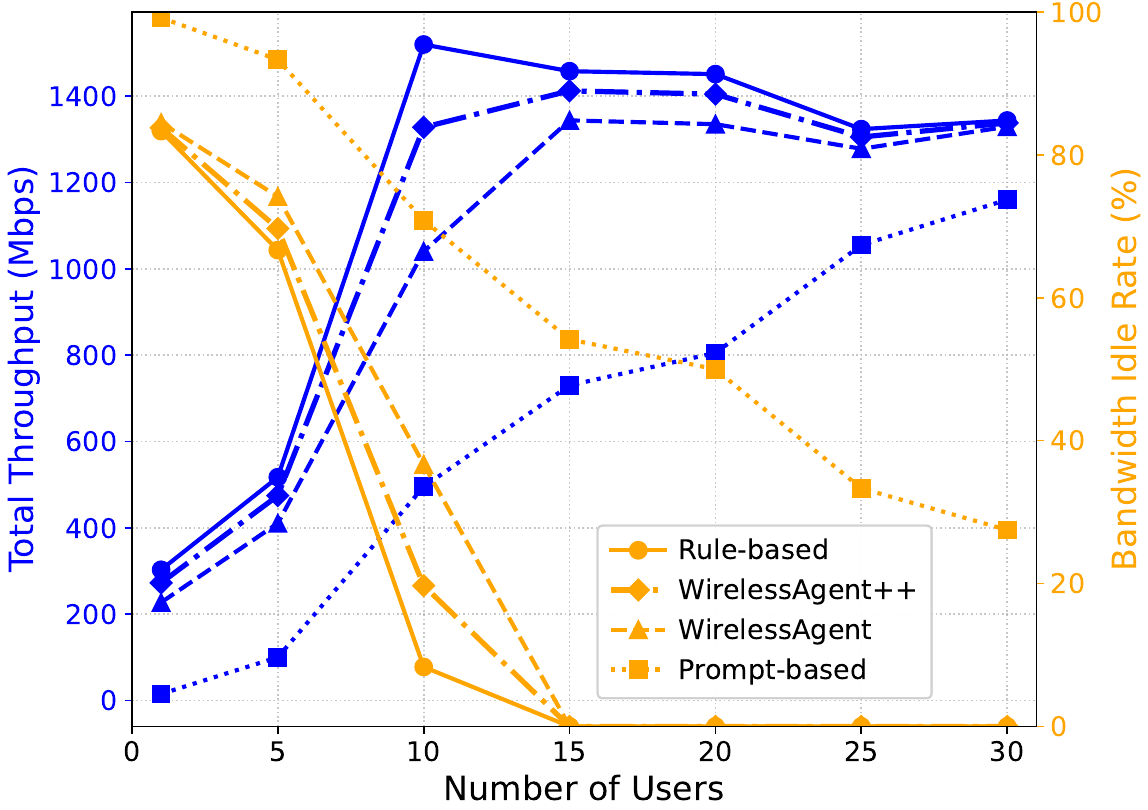}}
\caption{Total throughput and bandwidth idle rate versus the number of users under three HKUST campus scenarios~\cite{tong2024wirelessagent}. \wap{} consistently achieves the highest throughput among all LLM-based methods.}
\label{fig:wcns_throughput}
\end{figure*}

Fig.~\ref{fig:wcns} presents the WCNS results. As shown in Fig.~\ref{fig:wcns}(a), the Rule-based method achieves a perfect composite score of $1.000$ by directly applying optimal algorithmic bandwidth allocation, serving as the performance upper bound. \wap{} attains a composite score of $0.922$, significantly outperforming the Prompt-based baseline ($0.613$, $+$30.9~pp) and approaching the hand-designed WirelessAgent~($0.985$) \emph{without any manual workflow engineering}. The per-metric breakdown in Fig.~\ref{fig:wcns}(b) reveals that \wap{} performs competitively on all four sub-metrics: slice type classification~($96\%$), CQI prediction~($96\%$), bandwidth allocation~($91\%$), and throughput calculation~($89\%$). The largest gain over Prompt-based appears on CQI prediction~($96\%$ vs.\ $2\%$), where tool-assisted ray-tracing (initially discovered via \texttt{ToolAgent} and subsequently compiled into a deterministic \texttt{CodeLevel} call) replaces unreliable LLM estimation.

\textbf{Workflow evolution.} Fig.~\ref{fig:wcns}(c) shows a three-phase trajectory analogous to WCHW.
\emph{Phase~1---Seed ($61.3\%$):} A bare \texttt{Custom} call; CQI prediction is essentially random.
\emph{Phase~2---Tool Discovery ($90.5\%$, $+$29.2~pp):} The optimizer discovers the ray-tracing tool via \texttt{ToolAgent}, replacing the LLM's unreliable CQI ``guess'' with algorithmic computation---the single largest gain.
\emph{Phase~3---Tool Compilation ($92.18\%$, $+$1.7~pp):} The \texttt{ToolAgent} call is compiled into a deterministic \texttt{CodeLevelRayTracing} operator (Appendix~\ref{OW_WCNS}), and the prompt is enriched with step-by-step instructions, CQI-to-$\eta$ mapping, and worked examples.
Fig.~\ref{fig:wcns_throughput} further confirms that \wap{} achieves the highest throughput and lowest bandwidth idle rate among all LLM-based methods across three geographic scenarios.

\subsubsection{WCMSA (Multi-Tool Mobile Service Assurance)}
\begin{figure*}[!t]
\centering
\subfloat[Overall composite score comparison]{\label{fig:wcmsa_overall}
\includegraphics[width=0.58\columnwidth]{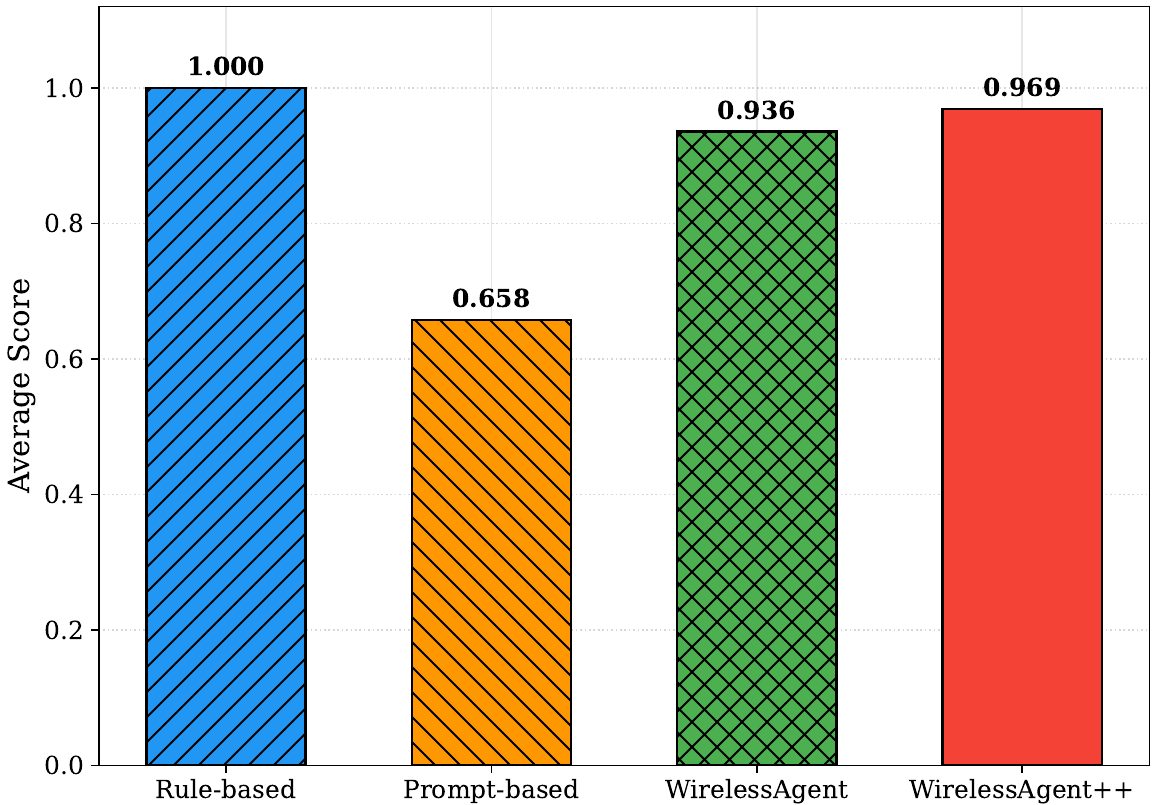}}
\hfil
\subfloat[Per-metric breakdown]{\label{fig:wcmsa_metric}
\includegraphics[width=0.83\columnwidth]{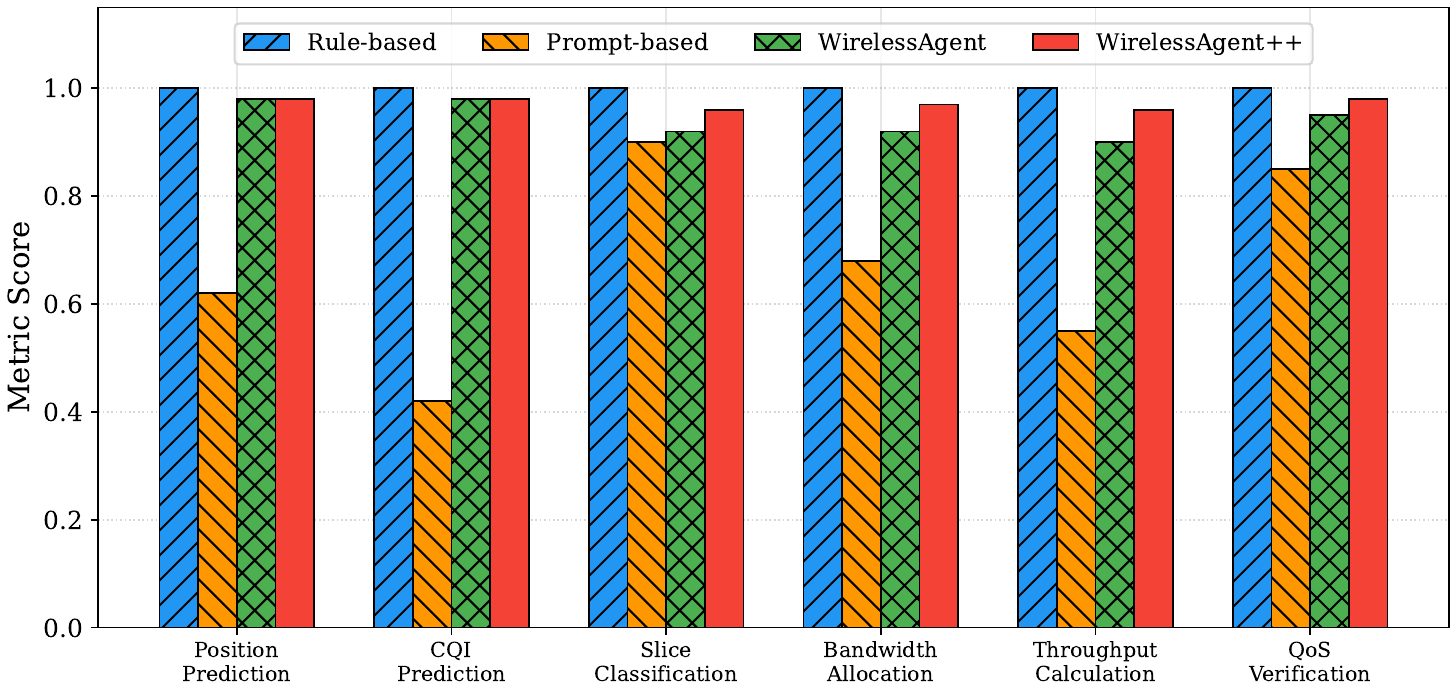}}
\hfil
\subfloat[Workflow evolution]{\label{fig:wcmsa_we}
\includegraphics[width=0.59\columnwidth]{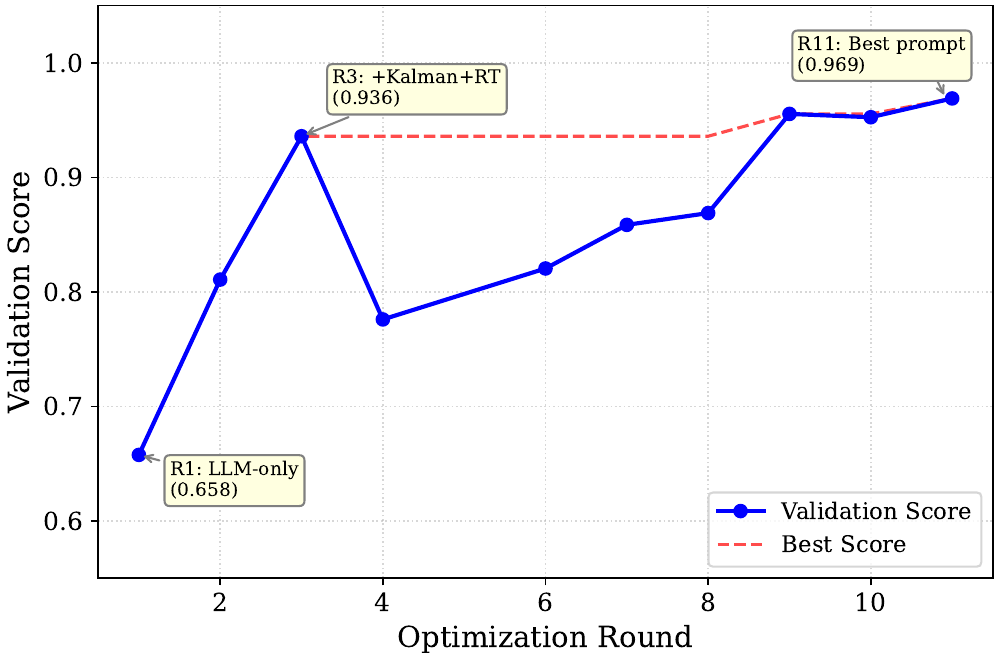}}
\caption{WCMSA case study. (a) Overall composite score comparison across methods. (b) Per-metric breakdown: \wap{} achieves $\geq$96\% on all six sub-metrics. (c) Workflow evolution on the validation set showing the three-phase optimization trajectory.}
\label{fig:wcmsa}
\end{figure*}

Fig.~\ref{fig:wcmsa} presents the WCMSA results. \wap{} achieves the highest composite score of $96.89\%$ and dominates all six sub-metrics: position prediction ($98\%$), CQI ($98\%$), slice type ($96\%$), bandwidth ($97\%$), throughput ($96\%$), and QoS satisfaction ($98\%$), as shown in Fig.~\ref{fig:wcmsa}(a)--(b). The improvement over WirelessAgent ($93.59\%$) is driven by the optimizer-discovered multi-tool pipeline, which chains the Kalman filter (position forecasting) and ray-tracing predictor (CQI estimation). Notably, the MCTS search initially discovers this pipeline via the \texttt{ToolAgent}'s ReAct loop and then \emph{compiles} it into deterministic \texttt{CodeLevel} operators (see Workflow Evolution below), enabling exact and cost-free tool execution.

\textbf{Workflow evolution.} Fig.~\ref{fig:wcmsa}(c) shows a three-phase trajectory.
\emph{Phase~1---Seed ($65.76\%$):} Without position prediction or channel estimation, accurate service assurance is impossible.
\emph{Phase~2---Multi-Tool Discovery ($93.59\%$, $+$27.83~pp):} The optimizer discovers a multi-tool ReAct chain: Kalman filter (position forecasting) $\to$ ray-tracing (CQI estimation) $\to$ \texttt{Custom} (reasoning). This \emph{predict-then-estimate-then-reason} pattern is unattainable with single-shot invocations.
\emph{Phase~3---Tool Compilation ($96.89\%$, $+$1.35~pp):} As in WCNS, the multi-tool chain is compiled into deterministic \texttt{CodeLevel} operators (Appendix~\ref{OW_WCMSA}), and the prompt is refined with a CQI-to-$\eta$ lookup table and a 7-point verification checklist.

\subsection{Ablation Study}
\label{sec:ablation}

We conduct ablation experiments on WCHW to study the contribution of each proposed component. Results are shown in Table~\ref{tab:ablation}. Removing domain tools (i.e., disabling \texttt{ToolAgent} and the formula retriever) incurs the largest degradation, reducing accuracy by 19.3 percentage points---consistent with the Round~1 (no tools, 62.44\%) vs.\ Round~2 (with tools, 80.86\%) transition observed during optimization. Disabling the 3-class experience mechanism (reverting to binary success/failure) causes the optimizer to make redundant mutations that chase noise, reducing convergence quality. The Penalized Boltzmann Selection prevents the optimizer from being trapped in poorly performing subtrees, and removing it leads to wasted exploration rounds. The Heuristic Critic provides modest but consistent gains by pre-screening mutations before expensive evaluation.

\begin{table}[t]
\centering
\caption{Ablation study on WCHW validation set. $\Delta$ denotes the change from the full \wap{} configuration.}
\label{tab:ablation}
\footnotesize
\begin{tabular}{@{}lcc@{}}
\toprule
\textbf{Configuration} & \textbf{Accuracy (\%)} & $\boldsymbol{\Delta}$ \\
\midrule
\wap{} (Full) & \textbf{81.78} & -- \\
\quad w/o Penalized Selection & 79.11 & $-2.67$ \\
\quad w/o Heuristic Critic & 80.34 & $-1.44$ \\
\quad w/o 3-Class Experience & 78.59 & $-3.19$ \\
\quad w/o Domain Tools & 62.44 & $-19.34$ \\
\midrule
AFlow (No enhancements)$^\dagger$ & 69.92 & $-11.86$ \\
\bottomrule
\end{tabular}
\end{table}

\subsubsection{Effect of Significance Threshold $\epsilon$}
The 3-class experience threshold $\epsilon$ controls the sensitivity of mutation outcome classification. With $\epsilon = 0.02$ (our default, corresponding to $2\%$ absolute accuracy), mutations that change the score by less than $2$ percentage points are classified as \textsc{Neutral} rather than Success/Failure. This prevents the optimizer from treating noise-induced fluctuations as meaningful signals. In the WCHW trajectory, $5$ out of $18$ mutations were classified as \textsc{Neutral} (score change $<2\%$), including the Round~$18$ mutation ($\Delta s = -0.016$) which would have been incorrectly classified as a failure under binary classification. Smaller values of $\epsilon$ (e.g., $0.005$) lead to more volatile optimization behavior, while larger values (e.g., $0.10$) cause the optimizer to miss genuine improvements.

\subsubsection{Search Cost Analysis}
Table~\ref{tab:search_cost} summarizes the MCTS search cost across all three benchmarks. The total optimization cost remains below \$5 for all tasks, and per-problem inference cost is under \$0.001. Notably, the WCNS and WCMSA searches complete in $\sim$13--14~minutes, since tool-calling workflows require fewer optimization rounds than knowledge-reasoning workflows.

\begin{table}[t]
\centering
\caption{MCTS search cost and inference cost across benchmarks.}
\label{tab:search_cost}
\footnotesize
\begin{tabular}{@{}lccc@{}}
\toprule
& \textbf{WCHW} & \textbf{WCNS} & \textbf{WCMSA} \\
\midrule
Search Rounds          & 19     & 11     & 11 \\
Wall-Clock Time (min)  & 63     & 13     & 14 \\
Total Search Cost (USD)& 4.95   & 0.99   & 1.05 \\
Per-Problem Cost (USD) & 0.00083 & 0.00056 & 0.00068 \\
\bottomrule
\end{tabular}
\end{table}

\subsection{Discussion}\label{sec:discussion}

\textbf{Workflow transferability across LLMs.}
The workflows discovered by \wap{} are optimized with Claude-Opus-4.5 as the optimizer and Qwen-turbo-latest as the executor. An important open question is whether these optimized workflows transfer to other executor LLMs (e.g., GPT-4o, Llama-3). While the structural components (operator graphs and tool-calling patterns) are LLM-agnostic, the prompt instructions are calibrated to the executor's specific strengths and formatting conventions. Investigating cross-model workflow transfer (and developing model-agnostic optimization strategies) is a promising direction for future work.

\textbf{Hyperparameter sensitivity.}
Our default hyperparameters (top-$K=5$, $\lambda=0.3$, $\alpha=0.2$, $\epsilon=0.02$, temperature$=0$) were empirically selected on the WCHW validation set (Section~\ref{sec:ablation}). The effect of $\epsilon$ is analyzed in detail above; we observe that the remaining hyperparameters are relatively stable across benchmarks, as the same settings yield strong performance on all three tasks without per-task tuning.

\textbf{Scoring metric design.}
The composite scoring metrics for WCNS and WCMSA involve sub-metrics on different scales---some in dB, others as ratios or percentages. We adopt a relative-error threshold of $\leq 5\%$ for numerical comparisons and weighted sub-metric aggregation. While this design is conservative and aligns with standard engineering tolerances, alternative metric designs (e.g., rank-based or percentile-based scoring) may yield different optimization dynamics. We release the full scoring code with our benchmark to facilitate community experimentation.

\textbf{Rule-based oracle.}
The Rule-based method serves as a performance upper bound in WCNS and WCMSA by applying optimal algorithmic allocation with \emph{perfect knowledge} of CQI values computed directly from path-loss models. In practice, such perfect channel knowledge is unavailable; the Rule-based method thus represents an idealized oracle rather than a deployable baseline. \wap{}'s performance gap relative to this oracle ($\sim$8~pp on WCNS, $\sim$3~pp on WCMSA) primarily stems from residual errors in CQI-conditioned reasoning rather than channel estimation.

\textbf{Wireless modeling simplifications.}
The current \wb{} benchmarks model a simplified wireless environment: single-cell ray-tracing with fixed infrastructure, no inter-cell interference, no MIMO spatial multiplexing, and ideal fronthaul. These simplifications are intentional to isolate the agent's reasoning capability from the complexity of the physical layer. Extending the benchmark to multi-cell interference management, beamforming optimization, and RIS-aided scenarios is a natural next step that would test more sophisticated tool-agent interactions.

\textbf{Limitations.}
Our framework has several limitations that warrant further investigation. First, \wap{} currently operates in a single-agent setting; extending to multi-agent collaboration (e.g., cooperative network management across base stations) is an important direction. Second, the current tool library is fixed during MCTS optimization; future work could allow the optimizer to dynamically discover or compose new tools. Third, while our three $\epsilon$-thresholds ($0.02$) and critic aggressiveness levels ($0.50$, $0.65$) work well across \wb{}, they may need recalibration for tasks with very different score distributions. Finally, the WCMSA benchmark uses a simplified mobility model (constant-velocity Kalman filter); more realistic mobility patterns (group mobility, sudden stops, indoor--outdoor transitions) would further stress-test the agent's proactive reasoning capabilities.

\section{Conclusions}
\label{sec:conclusion}
We presented \wap{}, a framework that automates the design of LLM-based autonomous agents for wireless network tasks. By representing agent workflows as executable programs and applying a domain-adapted MCTS algorithm with penalized Boltzmann selection, maturity-aware heuristic critic, and 3-class experience replay, \wap{} discovers high-performing workflows without manual engineering. Our \wb{} benchmark suite establishes a standardized evaluation framework covering knowledge reasoning (WCHW, 1,392 problems), code-based network slicing (WCNS, 1,000 problems), and mobile service assurance (WCMSA, 1,000 problems). Experiments demonstrate that \wap{} achieves $78.37\%$  on WCHW, $90.95\%$ on WCNS, and $97.07\%$ on WCMSA, significantly outperforming prompt-based baselines (by up to 31~pp) and general-purpose workflow optimizers (by up to 11.1~pp). This work marks an important step from ``building agents'' to ``building agent builders'' for next-generation wireless networks.


\bibliographystyle{IEEEtran}
\bibliography{WAP_Ref}

\appendix
\subsection{WCHW Scoring Pipeline}\label{app:scoring}

The WCHW evaluation employs a multi-strategy scoring pipeline to handle the diversity of answer formats in wireless communication problems. The pipeline proceeds as follows:

\textbf{Step~1: Answer format classification.} Each reference answer is classified into one of nine types using regex-based pattern matching: \textsc{PureNumeric}, \textsc{NumericWithUnit}, \textsc{Scientific} (e.g., $2.13 \times 10^{-2}$), \textsc{Formula} (e.g., $h(t) = A[U(t-2T/3) - U(t-T)]$), \textsc{Percentage}, \textsc{TextShort}, \textsc{TextLong}, \textsc{CodeSequence} (e.g., binary codewords), and \textsc{Ratio}.

\textbf{Step~2: Unit detection and normalization.} The system recognizes 38 unit multipliers spanning six physical dimensions: frequency (Hz, kHz, MHz, GHz), data rate (bps, kbps, Mbps, Gbps), power (W, mW, dBm, dBW), time (s, ms, $\mu$s, ns), distance (m, km), and spectral efficiency (bps/Hz). All extracted values are automatically converted to base SI units before comparison.

\textbf{Step~3: Multi-strategy scoring.} Four independent scoring strategies are applied in parallel, and the \emph{maximum} score is returned:
\begin{enumerate}[leftmargin=*]
    \item \textbf{Format-aware matcher:} Uses the classified format type to apply format-specific comparison (e.g., Hamming distance for code sequences, structural equivalence for formulas).
    \item \textbf{Numeric scorer:} Extracts numeric values (including scientific notation) and computes relative error with the tiered scheme in Table~\ref{tab:wchw_scoring}.
    \item \textbf{Formula scorer:} Normalizes LaTeX representations, checks exact match ($\to$ 1.0), containment ($\to$ 0.8), and Jaccard character-level similarity with variable overlap ($>0.7 \to 0.8$, $>0.5 \to 0.5$).
    \item \textbf{Text scorer:} Computes weighted keyword overlap (60\% number match + 40\% word match), with thresholds $>0.8 \to 1.0$, $>0.6 \to 0.8$, $>0.4 \to 0.5$.
\end{enumerate}
This max-of-four-strategies approach reduces false negatives caused by format mismatches between the agent's output and the reference answer.

\subsection{WCHW Dataset Examples}\label{Ex_WCHW}

WCHW problems span 10 knowledge categories with varying difficulty levels. Below are two representative examples:

\begin{tcolorbox}[breakable, colback=orange!5!white, colframe=orange!75!black, title={WCHW Example: Wireless Knowledge Reasoning}]
\footnotesize
\textbf{Question 1:} Shannon capacity. $B=50$ MHz, SNR$=0.1$ (linear). Compute $C$ (Mbps).

\textbf{Answer:} 6.87 Mbps

\textbf{CoT:} Step 1: $C=B\log_2(1+\text{SNR})=50\times10^6\log_2(1.1)$. Step 2: $\log_2(1.1)=0.1375$. Step 3: $C=50\times10^6 \times 0.1375=6.87\times10^6$ bps $=6.87$ Mbps.

\space{. . .}

\textbf{Question 20:} Compute BER for noncoherent BFSK at $E_b/N_0=8$ dB.

\textbf{Answer:} $2.13\times10^{-2}$

\textbf{CoT:} Step 1: Convert to linear: $\gamma=10^{0.8}=6.31$. Step 2: $P_b=0.5 e^{-\gamma/2}=0.5 e^{-3.155}$. Step 3: $e^{-3.155}=0.0426$. Step 4: $P_b=0.0213$.
\end{tcolorbox}

\subsection{WCNS Dataset Examples}\label{Ex_WCNS}

WCNS problems require service intent classification and resource allocation:

\begin{tcolorbox}[breakable, colback=blue!5!white, colframe=blue!60!black, title={WCNS Example: Network Slicing Decision}]
\footnotesize
\textbf{Network State:} eMBB Slice: 12 active users (90 MHz capacity); URLLC Slice: 3 active users (30 MHz capacity)

\textbf{New User:} CQI = 8, Service Request: ``I want to browse websites and check email''

\textbf{Answer:} Slice Type: eMBB, Bandwidth: 6.92 MHz, Throughput: 13.2 Mbps

\textbf{CoT:} (1) Intent: Web browsing is high-throughput, low-latency-tolerant $\rightarrow$ eMBB. (2) Bandwidth: $B=90/(12+1)=6.92$ MHz. (3) CQI=8 $\rightarrow$ $\eta=1.91$ bps/Hz (3GPP TS~38.214 Table~5.2.2.1-2). (4) Throughput: $R=B\cdot\eta=6.92 \times 1.91=13.2$ Mbps.
\end{tcolorbox}

\subsection{WCMSA Dataset Examples}\label{Ex_WCMSA}

WCMSA problems require trajectory prediction and proactive resource allocation:

\begin{tcolorbox}[breakable, colback=green!5!white, colframe=green!60!black, title={WCMSA Example: Mobile Service Assurance}]
\footnotesize
\textbf{Trajectory:} Historical positions: $(79.3, 46.0) \rightarrow (80.1, 45.4) \rightarrow (81.2, 44.7) \rightarrow (82.1, 44.1)$

\textbf{Base Station:} Location $(0, 0)$, Tx power: 30 dBm, Carrier: 2.4 GHz

\textbf{Service Request:} ``cloud gaming with high graphics'', Min rate: 35 Mbps

\textbf{Answer:} Predicted Position: $(83.0, 43.5)$, CQI: 15, Slice: eMBB, Bandwidth: 20 MHz, Throughput: 111.0 Mbps, QoS Satisfied: Yes

\textbf{CoT:} (1) Kalman filter prediction: $\hat{x}=82.1+(82.1-81.2)=83.0$, $\hat{y}=44.1+(44.1-44.7)=43.5$. (2) Distance to BS: $d=\sqrt{83^2+43.5^2}=93.7$ m. (3) Ray-tracing $\rightarrow$ SNR~(dB) $\rightarrow$ CQI=15. (4) CQI=15 $\rightarrow$ $\eta=5.55$ bps/Hz (3GPP TS~38.214 Table~5.2.2.1-2). (5) Cloud gaming is high-throughput $\rightarrow$ eMBB. (6) $B=\min(90/3, 20)=20$ MHz. (7) $R=B\cdot\eta=20 \times 5.55=111.0$ Mbps $>35$ Mbps $\checkmark$.
\end{tcolorbox}

\subsection{WCHW Optimal Workflow}\label{OW_WCHW}

The optimal WCHW workflow (Round~14, 81.78\%) implements a \emph{Reason-then-Verify} pattern via \textit{Custom}$\,\to\,$\textit{ToolAgent}:

\begin{tcolorbox}[breakable, colback=orange!5!white, colframe=orange!75!black, title={WCHW Optimal Workflow (Round 14, 81.78\%)}]
\scriptsize
\textbf{Input:} \texttt{problem} (telecom question string)\\[1pt]
\fbox{\textbf{Stage 1: \textit{Custom}}} $\longleftarrow$ \texttt{SOLVE\_PROMPT}\\[1pt]
\texttt{solution = Custom(input=problem, instruction=SOLVE\_PROMPT)}\\[1pt]
\textit{Custom} calls the Executor LLM with a domain-enriched prompt containing $>$10 critical formulas:
\begin{itemize}[leftmargin=*, nosep, topsep=1pt]
  \item Shannon: $C = B\log_2(1+\text{SNR}_{\text{linear}})$; Matched filter: $h(t)=s(T-t)$
  \item BER: BPSK $0.5\,\text{erfc}(\sqrt{E_b/N_0})$; BFSK $0.5\,e^{-E_b/(2N_0)}$; DPSK $0.5\,e^{-E_b/N_0}$
  \item FM Carson $\text{BW}=2(\Delta f + f_m)$; NOMA/SIC ordering; water-filling
  \item PCM SQNR $= 6.02n + 1.76$~dB; DM SNR; NRZ/raised-cosine BW
\end{itemize}
$\big\downarrow$ \texttt{solution['response']} (LLM-derived answer)\\[2pt]
\fbox{\textbf{Stage 2: \textit{ToolAgent}}} $\longleftarrow$ ReAct loop, max\_steps = 2\\[1pt]
\texttt{verified = ToolAgent(problem + solution, "Verify via Python code")}\\[1pt]
\textit{ToolAgent} generates Python code to \emph{independently recalculate} the answer (e.g., \texttt{math.log2()}, \texttt{math.erfc()}), compares against Stage~1, and emits the verified value.\\[2pt]
$\big\downarrow$\\[1pt]
\textbf{Output:} \texttt{verified['answer']} (final numerical answer in base units)
\end{tcolorbox}

\subsection{WCNS Optimal Workflow}\label{OW_WCNS}

The optimal WCNS workflow (Round~10, 92.18\%) implements a \emph{Tool-then-Reason} pattern via \textit{CodeLevelRayTracing}$\,\to\,$\textit{Custom}:

\begin{tcolorbox}[breakable, colback=blue!5!white, colframe=blue!60!black, title={WCNS Optimal Workflow (Round 10, 92.18\%)}]
\scriptsize
\textbf{Input:} \texttt{problem} (network slicing scenario with user position)\\[1pt]
\fbox{\textbf{Stage 1: \textit{CodeLevelRayTracing}}} (deterministic, no LLM)\\[1pt]
\texttt{cqi\_info = RayTracing.get\_cqi(problem)} \hfill\textit{\# parse $(x,y)$, region}\\\texttt{enhanced = RayTracing.inject\_cqi(problem, cqi\_info)}\\[1pt]
Extracts coordinates and region, queries the pre-trained ray-tracing channel model, returns exact CQI (1--15). Injects: \texttt{``ACCURATE CQI FROM RAY TRACING: \{cqi\}''}\\[2pt]
$\big\downarrow$ \texttt{enhanced\_problem} (original problem + injected CQI)\\[2pt]
\fbox{\textbf{Stage 2: \textit{Custom}}} $\longleftarrow$ \texttt{SOLVE\_PROMPT}\\[1pt]
\texttt{solution = Custom(input=enhanced, instruction=SOLVE\_PROMPT)}\\[1pt]
The prompt instructs the LLM to:
\begin{itemize}[leftmargin=*, nosep, topsep=1pt]
  \item Classify intent $\to$ eMBB or URLLC (keyword-based rules)
  \item Bandwidth: $B_{\text{eMBB}} = 90/(N\!+\!1)\in[6,20]$~MHz; $B_{\text{URLLC}} = 30/(N\!+\!1)\in[1,5]$~MHz
  \item Map CQI to spectral efficiency $\eta$ via 3GPP TS~38.214 Table~5.2.2.1-2; throughput: $R = B \cdot \eta$~(Mbps)
  \item Three worked examples: CQI$=5$ ($\eta\!=\!0.88$), $B\!=\!10{\to}8.8$; CQI$=10$ ($\eta\!=\!2.73$), $B\!=\!10{\to}27.3$; CQI$=15$ ($\eta\!=\!5.55$), $B\!=\!20{\to}111.0$~Mbps
\end{itemize}
$\big\downarrow$\\[1pt]
\textbf{Output:} \texttt{solution['response']} (CQI, Slice Type, Bandwidth, Throughput)
\end{tcolorbox}

\subsection{WCMSA Optimal Workflow}\label{OW_WCMSA}

The optimal WCMSA workflow (Round~11, 96.89\%) implements a \emph{Predict-Estimate-then-Reason} pattern via \textit{KalmanPredictor}$\,\to\,$\textit{RayTracing}$\,\to\,$\textit{Custom}:

\begin{tcolorbox}[breakable, colback=green!5!white, colframe=green!60!black, title={WCMSA Optimal Workflow (Round 11, 96.89\%)}]
\scriptsize
\textbf{Input:} \texttt{problem} (trajectory + network state + service request)\\[1pt]
\fbox{\textbf{Stage 1: \textit{CodeLevelKalmanPredictor}}} (deterministic, no LLM)\\[1pt]
\texttt{pos\_info = KalmanPredictor.predict(problem)}\\\texttt{enhanced = KalmanPredictor.inject\_prediction(problem, pos\_info)}\\[1pt]
Parses historical positions, fits a constant-velocity Kalman filter, injects: \texttt{``PREDICTED POSITION: (\{x\}, \{y\}), Region: \{region\}''}\\[2pt]
$\big\downarrow$ \texttt{enhanced\_problem} (problem + predicted position)\\[2pt]
\fbox{\textbf{Stage 2: \textit{CodeLevelRayTracing}}} (deterministic, no LLM)\\[1pt]
\texttt{cqi\_info = RayTracing.get\_cqi\_at(pos\_info.x, pos\_info.y, region)}\\\texttt{enriched = RayTracing.inject\_cqi(enhanced, cqi\_info)}\\[1pt]
Ray-tracing at the \emph{predicted} (not current) position---the key distinction from WCNS enabling \emph{proactive} allocation. Injects future CQI.\\[2pt]
$\big\downarrow$ \texttt{enriched\_problem} (problem + predicted position + future CQI)\\[2pt]
\fbox{\textbf{Stage 3: \textit{Custom}}} $\longleftarrow$ \texttt{SOLVE\_PROMPT}\\[1pt]
\texttt{solution = Custom(input=enriched, instruction=SOLVE\_PROMPT)}\\[1pt]
The prompt provides:
\begin{itemize}[leftmargin=*, nosep, topsep=1pt]
  \item eMBB/URLLC classification rules + user-count extraction
  \item Bandwidth formulas with clamping (same as WCNS)
  \item Precomputed CQI-to-$\eta$ lookup (3GPP TS~38.214 Table~5.2.2.1-2) for CQI$\in\{7,9,11,12,15\}$
  \item Four worked examples (e.g., $B\!=\!20$, CQI$\!=\!15$, $\eta\!=\!5.55$ $\to$ $R=20{\times}5.55=111.0$~Mbps)
  \item QoS check: Throughput $\geq$ min.\ rate $\to$ Yes/No; 7-point checklist
\end{itemize}
$\big\downarrow$\\[1pt]
\textbf{Output:} \texttt{solution['response']} (Position, CQI, Slice, BW, Throughput, QoS)
\end{tcolorbox}

\subsection{CQI-to-Spectral-Efficiency Mapping}\label{app:cqi_table}

Table~\ref{tab:cqi_full} provides the complete CQI-to-spectral-efficiency mapping used in \wb{}, derived from the 4-bit CQI Table~1 of 3GPP TS~38.214 (Table~5.2.2.1-2). This mapping is used by both the ray-tracing channel predictor (Section~\ref{sec:ray_tracing_tool}) and the LLM prompts in the optimized workflows.

\begin{table}[t]
\centering
\caption{Complete CQI-to-spectral-efficiency mapping (3GPP TS~38.214, Table~5.2.2.1-2).}
\label{tab:cqi_full}
\footnotesize
\begin{tabular}{@{}cccc@{}}
\toprule
\textbf{CQI} & \textbf{Modulation} & \textbf{Code Rate $\times 1024$} & $\boldsymbol{\eta}$ \textbf{(bps/Hz)} \\
\midrule
1  & QPSK    & 78   & 0.15 \\
2  & QPSK    & 120  & 0.23 \\
3  & QPSK    & 193  & 0.38 \\
4  & QPSK    & 308  & 0.60 \\
5  & QPSK    & 449  & 0.88 \\
6  & QPSK    & 602  & 1.18 \\
7  & 16-QAM  & 378  & 1.48 \\
8  & 16-QAM  & 490  & 1.91 \\
9  & 16-QAM  & 616  & 2.41 \\
10 & 64-QAM  & 466  & 2.73 \\
11 & 64-QAM  & 567  & 3.32 \\
12 & 64-QAM  & 666  & 3.90 \\
13 & 64-QAM  & 772  & 4.52 \\
14 & 64-QAM  & 873  & 5.12 \\
15 & 64-QAM  & 948  & 5.55 \\
\bottomrule
\end{tabular}
\end{table}

\subsection{Operator Repertoire}\label{app:operators}

Table~\ref{tab:operators} lists the complete set of operators available to the MCTS optimizer when constructing workflows. Each operator has a typed interface and specific computational characteristics.

\begin{table}[t]
\centering
\caption{Complete operator repertoire available to \wap{}.}
\label{tab:operators}
\footnotesize
\begin{tabular}{@{}lp{5.0cm}@{}}
\toprule
\textbf{Operator} & \textbf{Description} \\
\midrule
\texttt{Custom} & Invokes an LLM with input and instruction prompt; the most flexible operator \\
\texttt{ToolAgent} & ReAct-based agent (Algorithm~\ref{alg:react}) with tool calling; max $I$ iterations with early stopping \\
\texttt{CodeLevel} & Deterministic, LLM-free tool execution via compiled code (e.g., \texttt{CodeLevelRayTracing}, \texttt{CodeLevelKalmanPredictor}); zero variance, near-zero cost \\
\texttt{ScEnsemble} & Self-consistency ensemble~\cite{wang2023selfconsistency}: maps $N$ candidate solutions to letters and selects the most consistent answer \\
\texttt{MdEnsemble} & Medicine-inspired ensemble~\cite{nori2023medprompt}: shuffles $N$ answers, votes 5 times, returns majority answer \\
\texttt{Programmer} & Generates and executes Python code (3 retries, 30s timeout per attempt) \\
\texttt{Review} & Reviews a solution for correctness (returns boolean + feedback) \\
\texttt{Revise} & Revises a solution based on Review feedback \\
\texttt{Test} & Tests code against provided test cases (up to 3 reflection loops) \\
\texttt{AnswerGenerate} & Produces a structured thought $+$ answer pair in XML format \\
\texttt{Format} & Reformats solutions to match expected output schema \\
\texttt{AnswerValidator} & Rule-based $+$ LLM hybrid validation of answer format \\
\bottomrule
\end{tabular}
\end{table}

\subsection{WCMSA Service Types and QoS Requirements}\label{app:service_types}

Table~\ref{tab:service_types} lists the $20$ service types used in the WCMSA benchmark, each associated with a network slice type and minimum throughput requirement that the agent must verify.

\begin{table}[t]
\centering
\caption{WCMSA service types and minimum QoS requirements.}
\label{tab:service_types}
\footnotesize
\begin{tabular}{@{}llc@{}}
\toprule
\textbf{Service} & \textbf{Slice} & \textbf{Min Rate (Mbps)} \\
\midrule
4K video streaming          & eMBB  & 25 \\
HD movie download           & eMBB  & 50 \\
Cloud gaming (high graphics)& eMBB  & 35 \\
AR navigation (3D maps)     & eMBB  & 20 \\
VR social experience        & eMBB  & 40 \\
8K live sports broadcast    & eMBB  & 80 \\
Large file sync             & eMBB  & 30 \\
Holographic communication   & eMBB  & 60 \\
Remote desktop access       & eMBB  & 15 \\
Video analytics services    & eMBB  & 20 \\
\midrule
Remote robotic surgery      & URLLC & 10 \\
Autonomous vehicle V2X      & URLLC & 5 \\
Industrial IoT control      & URLLC & 2 \\
Real-time drone control     & URLLC & 8 \\
Emergency response coord.   & URLLC & 5 \\
Smart grid fault detection  & URLLC & 3 \\
Patient vital signs monitor & URLLC & 4 \\
High-frequency trading      & URLLC & 2 \\
Factory robot synchronization & URLLC & 6 \\
Precision CNC machine ctrl  & URLLC & 5 \\
\bottomrule
\end{tabular}
\end{table}

\subsection{Ray-Tracing Deployment Configuration}\label{app:ray_tracing_config}

The ray-tracing engine is deployed on three HKUST campus regions using OpenStreetMap building data. Table~\ref{tab:campus_regions} summarizes each region's characteristics. The transmitter is placed at the centroid of the tallest building in each region, elevated by 5.0~m above the rooftop. User positions are sampled uniformly from outdoor areas (verified outside building footprints via a ray-casting point-in-polygon algorithm). Each region provides a distinct propagation environment: the North region features dense low-rise academic buildings, the Center region has mixed-height structures around the main plaza, and the South region includes high-rise residential towers with deeper NLOS shadowing.

\begin{table}[t]
\centering
\caption{HKUST campus ray-tracing regions used in \wb{}.}
\label{tab:campus_regions}
\footnotesize
\begin{tabular}{@{}lccc@{}}
\toprule
\textbf{Region} & \textbf{OSM File} & \textbf{Typical Bldg.} & \textbf{Propagation} \\
\midrule
North  & \texttt{HKUST\_North.osm}  & Low-rise academic & LOS-dominant \\
Center & \texttt{HKUST\_Center.osm} & Mixed-height   & Moderate NLOS \\
South  & \texttt{HKUST\_South.osm}  & High-rise residential & Deep NLOS \\
\bottomrule
\end{tabular}
\end{table}

\subsection{MCTS Hyperparameter Summary}\label{app:hyperparams}

Table~\ref{tab:hyperparams} consolidates all hyperparameters used in \wap{} with their default values and descriptions.

\begin{table}[h]
\centering
\caption{Complete hyperparameter summary for \wap{}.}
\label{tab:hyperparams}
\footnotesize
\begin{tabular}{@{}lcp{3.4cm}@{}}
\toprule
\textbf{Parameter} & \textbf{Default} & \textbf{Description} \\
\midrule
$K$ (top-$K$)       & 5     & Parent candidates for Boltzmann selection \\
$\lambda$           & 0.3   & Uniform exploration weight in~\eqref{eq:boltzmann} \\
$\alpha$            & 0.2   & Boltzmann temperature in~\eqref{eq:boltzmann} \\
$\epsilon$          & 0.02  & 3-class significance threshold in~\eqref{eq:3class} \\
$V$                  & 5     & Validation runs per evaluation \\
$T$                  & 20    & Maximum optimization rounds \\
$k$ (convergence)    & 3     & Top-$k$ for convergence detection \\
$C$ (patience)       & 5     & Consecutive rounds for early stopping \\
$z$ (confidence)     & 0     & Significance level for convergence \\
$\tau_{\text{high}}$ & 0.65  & Critic: conservative threshold \\
$\tau_{\text{mid}}$  & 0.50  & Critic: moderate threshold \\
LLM temperature      & 0     & All models: deterministic decoding \\
\bottomrule
\end{tabular}
\end{table}

\subsection{Prompt for LLM Optimizer}\label{app:optimizer_prompt}

The MCTS optimizer uses the following system-level prompt to guide the Optimizer LLM (Claude-Opus-4.5) during the \emph{expansion} phase (Algorithm~\ref{alg:mcts}, Line~7). This prompt instructs the LLM to propose a single focused modification to the parent workflow.

\begin{tcolorbox}[breakable, colback=gray!5!white, colframe=gray!70!black, title={Workflow Optimize Prompt}]
\scriptsize
\texttt{WORKFLOW\_OPTIMIZE\_PROMPT =}\\[2pt]
\texttt{"""}You are building a Graph and corresponding Prompt to jointly solve \{type\} problems. Referring to the given graph and prompt, which forms a basic example of a \{type\} solution approach, please reconstruct and optimize them. You can add, modify, or delete nodes, parameters, or prompts. Include your single modification in XML tags in your reply. Ensure they are complete and correct to avoid runtime failures.

When optimizing, you can incorporate critical thinking methods like review, revise, ensemble (generating multiple answers through different/similar prompts, then voting/integrating/checking the majority to obtain a final answer), selfAsk, etc. Consider Python's loops, conditional statements, or machine learning techniques. The graph complexity should not exceed 10.

Output the modified graph and all the necessary Prompts in \texttt{prompt\_custom} (if needed). The prompt you need to generate is only the one used in \texttt{prompt\_custom.XXX} within Custom. Other methods already have built-in prompts and are prohibited from being generated. The generated prompt must not contain any placeholders.

Considering information loss, complex graphs may yield better results, but insufficient information transmission can omit the solution. It's crucial to include necessary context during the process.\texttt{"""}
\end{tcolorbox}

The optimizer receives the parent workflow's code, score, error logs, and experience record via a structured input template:

\begin{tcolorbox}[breakable, colback=gray!5!white, colframe=gray!70!black, title={Workflow Input Template (Expansion Phase)}]
\scriptsize
\texttt{WORKFLOW\_INPUT =}\\[2pt]
\texttt{"""}Here is a graph and prompt that performed excellently in a previous iteration (maximum score is 1). You must make further optimizations based on this graph. The modified graph must differ from the provided example.

\texttt{<sample>}\\
\quad\texttt{<experience>}\{experience\}\texttt{</experience>}\\
\quad\texttt{<score>}\{score\}\texttt{</score>}\\
\quad\texttt{<graph>}\{graph\}\texttt{</graph>}\\
\quad\texttt{<prompt>}\{prompt\}\texttt{</prompt>}\\
\quad\texttt{<operator\_description>}\{operator\_description\}\texttt{</operator\_description>}\\
\texttt{</sample>}

Below are the logs of results that encountered errors, which can be used as references for optimization: \{log\}

\textbf{SCORE-BASED OPTIMIZATION RULES} (YOUR CURRENT SCORE IS \{score\}):
\begin{itemize}[leftmargin=*, nosep, topsep=2pt]
  \item Score $\geq 0.65$ (HIGH): CONSERVATIVE --- Minor prompt tweaks only.
  \item Score 0.50--0.65 (MEDIUM): MODERATE --- Single structural change.
  \item Score $< 0.50$ (LOW): AGGRESSIVE --- Major restructuring allowed.
\end{itemize}
Only one detail point can be modified at a time, and no more than 5 lines of code may be changed per modification.\texttt{"""}
\end{tcolorbox}

\subsection{Operator Prompts}\label{app:operator_prompts}

Below are the core prompt templates used by the operators in $\mathcal{O}$. Each prompt is passed to the Executor LLM via a Pydantic-constrained output schema.

\begin{tcolorbox}[breakable, colback=orange!5!white, colframe=orange!75!black, title={ScEnsemble Prompt (Self-Consistency Voting)}]
\scriptsize
\texttt{SC\_ENSEMBLE\_PROMPT =}\\[2pt]
\texttt{"""}Given the question: \{question\}\\
Several solutions have been generated: \{solutions\}

Carefully evaluate these solutions and identify the answer that appears most frequently across them. This consistency in answers is crucial for determining the most reliable solution.

In the ``thought'' field, provide a detailed explanation of your thought process. In the ``solution\_letter'' field, output only the single letter ID (A, B, C, etc.) corresponding to the most consistent solution.\texttt{"""}
\end{tcolorbox}

\begin{tcolorbox}[breakable, colback=orange!5!white, colframe=orange!75!black, title={Review Prompt}]
\scriptsize
\texttt{REVIEW\_PROMPT =}\\[2pt]
\texttt{"""}Given a problem and a solution, review the solution's correctness using critical thinking and provide a result in boolean format.

problem: \{problem\}\quad solution: \{solution\}

If you are more than 95 percent confident that the final answer is incorrect, return False and give feedback for the error. Otherwise, return True and explain the correctness.\texttt{"""}
\end{tcolorbox}

\begin{tcolorbox}[breakable, colback=orange!5!white, colframe=orange!75!black, title={Revise Prompt}]
\scriptsize
\texttt{REVISE\_PROMPT =}\\[2pt]
\texttt{"""}Given a problem and a solution which is just reviewed as incorrect, revise the solution to solve the question.

problem: \{problem\}\quad solution: \{solution\}\quad feedback: \{feedback\}

Ensure the output is self-contained, without additional text or test cases.\texttt{"""}
\end{tcolorbox}

\subsection{ReAct Agent Protocol}\label{app:react_prompt}

The \texttt{ToolAgent} operator (Algorithm~\ref{alg:react}) uses a three-layer prompt architecture. The fixed protocol layer enforces structured XML output:

\begin{tcolorbox}[breakable, colback=blue!5!white, colframe=blue!60!black, title={ReAct Agent Fixed Protocol}]
\scriptsize
\texttt{REACT\_AGENT\_FIXED\_PROTOCOL =}\\[2pt]
\texttt{"""}Your response MUST be in valid XML format with ALL required tags.

If you want to use a tool, output EXACTLY this format:\\
\texttt{<thought>}Your reasoning about what to do next\texttt{</thought>}\\
\texttt{<action\_type>}use\_tool\texttt{</action\_type>}\\
\texttt{<tool\_name>}name of the tool\texttt{</tool\_name>}\\
\texttt{<tool\_args>}\{``arg1'': ``value1'', ``arg2'': ``value2''\}\texttt{</tool\_args>}

If you want to provide the final answer, output EXACTLY this format:\\
\texttt{<thought>}Your reasoning about why you can answer now\texttt{</thought>}\\
\texttt{<action\_type>}final\_answer\texttt{</action\_type>}\\
\texttt{<final\_answer>}Your CONCISE answer to the question\texttt{</final\_answer>}

STRICT REQUIREMENTS:
\begin{itemize}[leftmargin=*, nosep, topsep=1pt]
  \item ALWAYS include \texttt{<thought>} and \texttt{<action\_type>} tags.
  \item \texttt{tool\_args} MUST be valid JSON: double quotes, no trailing commas.
  \item All numeric values must be pre-computed numbers, not expressions.
  \item \texttt{<final\_answer>} tag MUST contain the actual answer, NOT be empty.
\end{itemize}\texttt{"""}
\end{tcolorbox}

The strategy prompt (optimizable by MCTS) and runtime context are prepended:
\begin{tcolorbox}[breakable, colback=blue!5!white, colframe=blue!60!black, title={ReAct Assembled Prompt Structure}]
\scriptsize
Each iteration of the ReAct loop sends the following to the Executor LLM:
\begin{verbatim}
{strategy_prompt}         # Layer 1: Optimizable (MCTS)

Available Tools:
{tools_description}       # Auto-generated from tool registry

Question: {problem}

Previous Steps:
Step 1: thought=... action=... result=...
Step 2: thought=... action=... result=...

{REACT_AGENT_FIXED_PROTOCOL}  # Layer 3: Immutable
\end{verbatim}
\end{tcolorbox}

\subsection{Heuristic Critic Report}\label{app:critic_report}

The heuristic critic (Section~\ref{sec:critic}) generates a structured report that is appended to the optimizer prompt. Below is a representative example:

\begin{tcolorbox}[breakable, colback=red!5!white, colframe=red!60!black, title={Heuristic Critic Report Example (Conservative Mode)}]
\scriptsize
\begin{verbatim}
=== SMART CRITIC REPORT (Round 14) ===
Score: 0.8178 | Mode: CONSERVATIVE

[ERROR ANALYSIS]
  Total Errors: 63
  - Format/Extraction: 22 (35%)
  - Unit Mismatch: 6 (9%)
  - Calculation/Value: 35 (56%)

[WORKFLOW COMPLEXITY]
  Steps: 2 | ToolAgent: Yes | ScEnsemble: No
  Conditionals: No | Custom Calls: 1

[EXPLORATION HISTORY]
  Attempts: 5 | Success: 2 | Failure: 2
  Node Saturated: No

[RECOMMENDATIONS]
  HIGH SCORE - BE VERY CAREFUL!
  [ALLOWED]
  - Minor prompt wording improvements
  - Add/remove one instruction line
  - Clarify unit conversion instructions
  [FORBIDDEN]
  - Adding new operators (ToolAgent, ScEnsemble)
  - Removing existing operators
  - Changing workflow structure
  - Adding if/else conditional logic
\end{verbatim}
\end{tcolorbox}

\subsection{3-Class Experience Replay Format}\label{app:experience_format}

The experience record provided to the optimizer during expansion is formatted as follows:

\begin{tcolorbox}[breakable, colback=green!5!white, colframe=green!60!black, title={Formatted Experience (Tree Structure)}]
\scriptsize
\begin{verbatim}
{
  "1": {
    "score": 0.6244,
    "success": {
      "2": {
        "modification": "Add ToolAgent operator to
          verify numerical calculations via Python",
        "score": 0.8086
      }
    },
    "failure": {
      "8": {
        "modification": "Add ScEnsemble operator
          to generate multiple solutions and vote",
        "score": 0.4336
      }
    }
  },
  "2": {
    "score": 0.8086,
    "success": {},
    "failure": {},
    "neutral": {
      "14": {
        "modification": "Expand formula library
          with water-filling, NOMA/SIC, and
          matched-filter formulas",
        "score": 0.8178
      },
      "18": {
        "modification": "Reword ToolAgent prompt
          for formula-type answers",
        "score": 0.8014
      }
    }
  }
}
\end{verbatim}
The tree records the parent--child relationship: the root node (Round~1, score 0.6244) has a successful child (Round~2, $+$18.4~pp via ToolAgent addition) and a failed child (Round~8, ScEnsemble caused a $-$19.1~pp drop). Round~2 in turn has two neutral children (Rounds~14 and~18), both with $|\Delta s| < \epsilon = 0.02$: Round~14 ($+$0.92~pp, formula expansion) and Round~18 ($-$0.72~pp, prompt rewording). This illustrates that even a best-round improvement can be classified as \textsc{Neutral} under the $\epsilon$-threshold, preventing the optimizer from over-fitting to noise.
\end{tcolorbox}

\end{document}